\documentclass[final,letterpaper]{IEEEtran}
\usepackage{color}
\usepackage{amsmath}
\usepackage{amsthm}
\usepackage{amssymb}
\usepackage{graphicx}
\usepackage{mathtools}
\allowdisplaybreaks

\usepackage{url}
\usepackage{cite}
\usepackage{epsfig}
\usepackage{algorithm}
\usepackage{algorithmic}
\usepackage{comment}

\DeclareMathOperator*{\argmin}{\arg\min}
\DeclareMathOperator*{\argmax}{\arg\max}

\makeatletter

\newcommand{\lyxmathsym}[1]{\ifmmode\begingroup\def\b@ld{bold}
  \text{\ifx\math@version\b@ld\bfseries\fi#1}\endgroup\else#1\fi}

\theoremstyle{plain}

\theoremstyle{plain}
\newtheorem{prop}{\protect\propositionname}
\theoremstyle{plain}
\newtheorem{cor}{\protect\corollaryname}
\theoremstyle{plain}
\newtheorem{lem}{\protect\lemmaname}
\theoremstyle{remark}
\newtheorem{rem}[]{\protect\remarkname}

\makeatother

\providecommand{\remarkname}{Remark}
\providecommand{\lemmaname}{Lemma}
\providecommand{\corollaryname}{Corollary}
\providecommand{\propositionname}{Proposition}
\providecommand{\theoremname}{Theorem}

\ifCLASSOPTIONcompsoc
\usepackage[caption=false,font=normalsize,labelfont=sf,textfont=sf,labelformat=simple]{subfig}
\else
\usepackage[caption=false,font=footnotesize,labelformat=simple]{subfig}
\fi

\begin{document}

\title{Cache-Aided Non-Orthogonal Multiple Access: The Two-User Case}
\author{Lin Xiang,~\IEEEmembership{Member,~IEEE,} Derrick Wing Kwan Ng,~\IEEEmembership{Senior~Member,~IEEE,} Xiaohu Ge,~\IEEEmembership{Senior~Member,~IEEE,} 
Zhiguo Ding,~\IEEEmembership{Senior~Member,~IEEE,} Vincent W.S. Wong,~\IEEEmembership{Fellow,~IEEE,} and Robert Schober,~\IEEEmembership{Fellow,~IEEE}

\thanks{This work has been presented in part at the \emph{IEEE International Conference on Communications (ICC)}, Kansas City, MO, May 2018 \cite{Xiang:ICC18:NOMA} and in the Ph.D. thesis of the first author \cite{Xiang:thesis18}. 
}
}

\maketitle

\begin{abstract}

In this paper, we propose a cache-aided non-orthogonal multiple access (NOMA) scheme for spectrally efficient downlink transmission. The proposed scheme not only reaps the benefits associated with NOMA and caching, but also exploits the data cached at the users for interference cancellation. As a consequence, caching can help to reduce the residual interference power, making multiple decoding orders at the users feasible. The resulting flexibility in decoding can be exploited for improved NOMA detection. We characterize the achievable rate region of cache-aided NOMA and derive the Pareto optimal rate tuples forming the boundary of the rate region. Moreover, we optimize cache-aided NOMA for minimization of the time required for completing file delivery. The optimal decoding order and the optimal transmit power and rate allocation are derived as functions of the cache status, the file sizes, and the channel conditions. Simulation results confirm that, compared to several baseline schemes, the proposed cache-aided NOMA scheme significantly expands the achievable rate region and increases the sum rate for downlink transmission, which translates into substantially reduced file delivery times.
\end{abstract}
\begin{IEEEkeywords}
Non-orthogonal multiple access (NOMA), Caching, Achievable rate region, Radio resource allocation, Convex and Nonconvex optimization, Pareto optimality
\end{IEEEkeywords}

\section{Introduction}

\IEEEPARstart{N}{on-orthogonal} multiple access (NOMA) is a key enabler for spectrally efficient wireless communications in the fifth-generation (5G) cellular networks \cite{SaitoVTCs13:NOMA,Ding17NOMA:MCOM,Ding17:JSAC:Survey}. NOMA pairs multiple simultaneous downlink transmissions on the same time-frequency resource via power domain or code domain multiplexing \cite{3GPP:TR36859}. By employing successive interference cancellation (SIC), strong user equipments (UEs) experiencing favorable channel conditions can cancel the interference caused by weak UEs suffering from poor channel conditions that are paired on the same time-frequency resource \cite{Tse2005Fundamentals}. Hence, strong UEs can achieve high data rates with low transmit power. As a consequence, the interference caused by strong UEs to weak UEs is limited and, at the same time, a high transmit power can be allocated to weak UEs to enhance user fairness \cite{Ding16UserP:TVT}. NOMA has also been extended to multicarrier and multi-antenna systems in \cite{Sun16FD-MC-NOMA:TCOM,Ding16MIMO-NOMA:TWC,George16MM:TSP}.

However, despite the growing interest in NOMA, the following limitations exist. First, as information-theoretic studies have shown, compared to conventional orthogonal multiple access (OMA), NOMA cannot increase the sum capacity, i.e., the maximum throughput of the system \cite{Tse2005Fundamentals,Cover12IT-Elements,Gamal11NIT}. For this reason, it is widely accepted that NOMA cannot meet the large capacity demands in 5G introduced by e.g. video streaming applications. Hence, NOMA has been mainly exploited to improve user fairness \cite{SaitoVTCs13:NOMA,Ding17NOMA:MCOM,Ding17:JSAC:Survey,Ding16UserP:TVT,Sun16FD-MC-NOMA:TCOM,Ding16MIMO-NOMA:TWC,George16MM:TSP}. Second, the performance gains of NOMA over conventional OMA are fundamentally limited by the users' channel conditions \cite{Ding16UserP:TVT}. For example, it is shown in \cite{Ding16UserP:TVT} that fixed-power NOMA can achieve a significant performance gain only when the channel gains of the UEs are substantially different.

For capacity enhancement in 5G networks, wireless caching has been recently proposed \cite{Schober5G}. Different from OMA and NOMA, caching exploits the statistical correlations of the source files, e.g., the video files intended for streaming, requested by the users \cite{Wang14:Cache}. By pre-storing the most popular video files in close proximity of the UEs, e.g. at base stations (BSs) and access points (APs), caching enables fast access to these files for UEs within the coverage of the BSs and the APs without burdening the backhaul links \cite{Ge16MCOM:UDN}. Moreover, advanced caching schemes introduce additional degrees of freedom which can be exploited in the physical layer to improve the quality of service (QoS) \cite{Xiang17TVT:CLCaching}, energy efficiency \cite{LiuJSAC16:EE}, and communication secrecy \cite{Xiang16TWC:CoMP,Xiang17TWC:Untrusted} during the delivery of cached and/or uncached contents without requiring additional spectral resources.

Recently, caching at streaming UEs, e.g. smartphones and tablets, has been advocated \cite{Walid17:Caching:mmWave,Ji:TIT16:D2D}, which can enhance the streaming quality of experience (QoE) while reducing (i.e., \emph{offloading}) over-the-air traffic just by utilizing the storage capacities available at the UEs. For example, UE side caching for enhanced mobility management of streaming UEs has been reported in \cite{Walid17:Caching:mmWave}. Since only minimal upgrade of the cellular infrastructure is required, UE side caching has been also proposed for opportunistic sharing of video files among UEs \cite{Ji:TIT16:D2D}. However, different from the caching at BSs and APs, caching at UEs poses significant challenges for the design of cache placement and delivery as the actual requests of each UE are difficult to predict during cache placement due to the users' mobility and the random nature of the users' requests. For example, files cached at a UE may not be requested by that UE later on, which reduces the performance gains enabled by caching.

In this paper, we propose the joint design of wireless caching and NOMA, which we refer to as cache-aided NOMA, to maximize the performance gains introduced by caching at the UEs. In fact, NOMA has been employed for instantaneous cache placement in a relatively straightforward manner in \cite{Ding17Noma:Caching}, where the placement and the delivery of video files are performed on the same time-frequency resource using conventional NOMA. However, due to the aforementioned capacity limitation of NOMA, the overhead introduced by instantaneous cache placement is non-negligible. Different from \cite{Ding17Noma:Caching}, the proposed cache-aided NOMA  exploits UE side caching to improve both the achievable rate region and the achievable sum rate of NOMA. 
The proposed scheme is inspired by {coded caching}, also referred to as cache-enabled coded multicast, as proposed in \cite{Niesen14IT:CodedCaching,Maddah:ToN15:decentralized}. In coded caching, exploiting the cached data as side information, a {coded multicast} format is created for simultaneous error-free video delivery to multiple users \cite{Niesen14IT:CodedCaching}. Appealingly, coded caching facilitates a multiplicative performance gain that scales with the aggregate cache memory size of the UEs even if there is no coordination for cache placement at the UEs \cite{Maddah:ToN15:decentralized}. Since the publication of \cite{Niesen14IT:CodedCaching,Maddah:ToN15:decentralized}, coded caching has attracted significant interest, see \cite{Tulino:MCOM17:Coding} and references therein. However, the caching concepts proposed in \cite{Niesen14IT:CodedCaching,Maddah:ToN15:decentralized,Tulino:MCOM17:Coding} are mainly applicable to noiseless and error-free communication links, as found e.g. in wireline networks. For wireless networks impaired by fading and noise, however, the performance of coded multicast is limited by the UE experiencing the worst channel conditions in the multicast group. To alleviate the adverse effects of fading, the authors of \cite{Timo:ISWCS15:Joint,Shirin17CacheAssign,Petros:ISIT17,Tulino:JSAC16:Channel-aware} have explored advanced {wireless coded caching} schemes employing {joint cache and channel coding} over erasure, degraded, and Gaussian broadcast channels. However, these advanced caching schemes rely on highly sophisticated network coding schemes for splitting files and forming multicast groups, which hinders their practical implementation.

Different from coded caching, the proposed cache-aided NOMA splits the video files into several subfiles and employs superposition coding of the requested uncached subfiles during delivery\footnote{As a result, unlike coded caching \cite{Niesen14IT:CodedCaching,Maddah:ToN15:decentralized,Tulino:MCOM17:Coding,Timo:ISWCS15:Joint,Shirin17CacheAssign,Petros:ISIT17,Tulino:JSAC16:Channel-aware}, cache-aided NOMA is directly applicable for arbitrary file and cache sizes.}. This allows the proposed cache-aided NOMA to jointly exploit cached and previously decoded data for improved interference cancellation at the UEs. In particular, if the cached content is a \emph{hit}, i.e., requested by the caching UE, cache-aided NOMA enables the conventional offloading of the video files. Furthermore, the missed cached data, which is not requested by the caching UE, is exploited by cache-aided NOMA as side information to facilitate (partial) interference cancellation. This cache-enabled interference cancellation (CIC) improves the signal-to-interference-plus-noise ratios (SINRs) of the received signals at both strong and weak UEs irrespective of their respective channel conditions. As a result, CIC may enable SIC even at weak UEs. By exploiting joint CIC and SIC at both strong and weak UEs, cache-aided NOMA can considerably reduce the impact of fading, which leads to improved user fairness and higher achievable sum rates compared to conventional NOMA. Furthermore, the number of feasible decoding orders increases compared to conventional NOMA. The decoding order can be optimized according to the cache and channel statuses for efficient resource allocation. The main contributions of this paper can be summarized as follows:
\begin{itemize}
\item We propose a novel cache-aided NOMA scheme which exploits cached data for cancellation of NOMA interference for spectrally efficient downlink file delivery. We characterize the achievable rate region of the proposed scheme and show that the Pareto optimal boundary rate tuples can be achieved by solving a rate maximization problem.

\item Based on the derived achievable rate regions for different decoding orders, we jointly  optimize the NOMA decoding order  and the transmit power and rate allocations for minimization of the file delivery time to enable \emph{fast} video delivery. Based on the optimality conditions, we obtain the optimal resource allocation in closed form. 

\item We show by simulation that the proposed scheme leads to a considerably larger achievable rate region, a significantly higher achievable sum rate, and a substantially lower file delivery time compared to several baseline schemes, including the straightforward combination of caching and NOMA.  
\end{itemize}

To the best of the authors' knowledge, this is the first work that exploits cached data for cancellation of NOMA interference. Hence, to provide insight into the design and performance of the proposed novel cache-aided NOMA scheme, we consider a two-user system. Nevertheless, an extension to more than two users is possible as will be explained in Section~\ref{sec2-4}. 

The remainder of this paper is organized as follows. In Section~\ref{sec2},  we present the system model and introduce the proposed cache-aided NOMA scheme. In Section~\ref{sec3}, the achievable rate regions of the proposed cache-aided NOMA scheme are characterized for different cache configurations and different decoding orders. The optimal power and rate allocation for delivery time minimization is investigated in Section~\ref{sec4}. The performance of the proposed scheme is evaluated in Section~\ref{sec5}, and finally, Section~\ref{sec6} concludes the paper. 

\emph{Notation:} $\mathbb{N}$, $\mathbb{C}$, and $\mathbb{R}_{+}$ denote the sets of natural, complex, and nonnegative real numbers, respectively. $\left|\mathcal{X}\right|$ denotes the cardinality of set $\mathcal{X}$. $\mathcal{CN}\left({\mu},{\sigma}^2\right)$ represents the complex Gaussian distribution with mean ${\mu}$ and variance ${\sigma}^2$. $\mathbb{E}(\cdot)$ and $(\cdot)^{T}$ are the expectation and the transpose operators, respectively. $\mathbf{1}\left[\cdot\right]$ denotes the indicator function which is $1$ if the event is true and $0$ otherwise. $\mathbf{x} \succeq \mathbf{y}$ ($\mathbf{x} \preceq \mathbf{y}$) means that vector $\mathbf{x}$ is element-wise greater (smaller) than or equal to vector $\mathbf{y}$. For decoding the received signals, the notation $i\overset{_{(m)}}{\to}x_{f}$ means that $x_{f}$ is the $m$th decoded signal at UE $i$. Similarly, $i\overset{_{(m)}}{\to}(x_{f},\,x_{f'})$ means that  signals $x_{f}$ and $x_{f'}$ are jointly decoded in the $m$th decoding step. Finally, $C( \Gamma )\triangleq \log_{2}\left(1+\Gamma\right)$ denotes the capacity function of an additive white Gaussian noise (AWGN) channel, where $\Gamma$ is the SINR.

\section{\label{sec2}System Model}

We consider cellular video streaming from a BS to two UEs indexed by $i$ and $j$, respectively. The BS and the UEs are single-antenna devices\footnote{{In small cells and Internet of Things (IoT) systems, deploying multiple antennas at the BSs and the UEs may not be possible due to stringent constraints on hardware complexity, energy consumption, and cost.}}. UE $k \in \{i,j\}$ is equipped with a cache and the sizes of the cache may vary across UEs. We consider the delivery of two files, $W_{A}$ and $W_{B}$, $W_{A}\neq W_{B}$, of sizes $V_{A}$ and $V_{B}$ bits, respectively. The system employs two transmission phases: the caching phase and the delivery phase. Assume that files $W_{A}$ and $W_{B}$ consist of sequentially organized video chunks. In the caching phase, UE $k\in\left\{ i,j\right\} $ places portions of the video chunks of file $W_f$, $f\in \{ A,B \}$, into its cache prior to the time of request, e.g. during the early mornings when  cellular traffic is low\footnote{We note that each UE may have cached multiple files but only the files requested by the considered UEs are relevant during delivery. Hence, only these files are considered here.}. We assume that UE $k$, $k\in\{i,j\}$, has cached $c_{kf}\in[0,1]$ portion of file $W_{f}$, $f\in \{ A,B \}$. {{In this paper, we focus on the delivery phase and assume that the values of $c_{kf}$ are given and fixed. The proposed delivery scheme is applicable for any caching scheme.}} In the delivery phase, the users' requests are known and the cached content is exploited for improved delivery of the requested files. Without loss of generality, we assume that UEs $i$ and $j$ request files $W_{A}$ and $W_{B}$, respectively. The requests are denoted as $(i,A)$ and $(j,B)$. We note that as the cache placement is completed before the users' requests are known, the UEs may cache files which the users later do not request. 

For ease of illustration, we start by assuming that the video chunks of $W_{A}$ and $W_{B}$ are cached in sequential order at the UEs. In this case, the video chunks of the same file cached at both UEs overlap. However, this assumption is not critical for the proposed cache-aided NOMA scheme and can be eliminated in a straightforward manner, as will be discussed in Section~\ref{sec2-4}.  

\subsection{Cache Status and File Splitting}

Let us define the minimum and the maximum cached portions of file $W_f$ as  
$\underline{c}_{f} \triangleq \min_{k\in\left\{ i,j\right\} } c_{kf}$ and $\overline{c}_{f} \triangleq \max_{k\in\left\{ i,j\right\} } c_{kf}$, which correspond to the cache status at UE  
$\underline{k}_{f}\triangleq\argmin_{k\in\left\{ i,j\right\} } c_{kf}$ and $\overline{k}_{f}\triangleq\argmax_{k\in\left\{ i,j\right\} } c_{kf}$, respectively\footnote{If $c_{if} = c_{jf}$, then we set $\underline{c}_{f} = c_{kf}$ and $\underline{k}_{f} = k$, $\forall (k,f) \in \{(i,A), (j,B)\}$, for convenience.}. Consequently, four UE cache configurations are possible at the time of request: 
\begin{itemize}
\item Case \mbox{I}: $i=\overline{k}_{B}$ and $j=\overline{k}_{A}$, i.e., $i=\underline{k}_{A}$ and $j=\underline{k}_{B}$. 

\item Case \mbox{II}: $i=\overline{k}_{B}$ and $j=\underline{k}_{A}$, i.e., $i=\overline{k}_{A}$ and $j=\underline{k}_{B}$.

\item Case \mbox{III}: $i=\underline{k}_{B}$ and $j=\overline{k}_{A}$, i.e., $i=\underline{k}_{A}$ and $j=\overline{k}_{B}$.

\item Case \mbox{IV}: $i=\underline{k}_{B}$ and $j=\underline{k}_{A}$, i.e., $i=\overline{k}_{A}$ and $j=\overline{k}_{B}$.

\end{itemize}
In general, Case \mbox{I} corresponds to the scenario where each UE has cached a smaller portion of the file that it is requesting than the other UE. This constitutes an unfavorable cache placement for both UEs but cannot be avoided in practice as the user requests cannot be accurately predicted. Similarly, Cases \mbox{II} and \mbox{III} correspond to unfavorable cache placements for UE $j$ and UE $i$, respectively. Finally, Case \mbox{IV} corresponds to favorable cache placements for both UEs, whereby the cached portions of the files are larger at the requesting UE than at the non-requesting UE.

\begin{figure}[t]
\centering
\subfloat[]{\label{fig1a} \centering\includegraphics[width=2.8in]{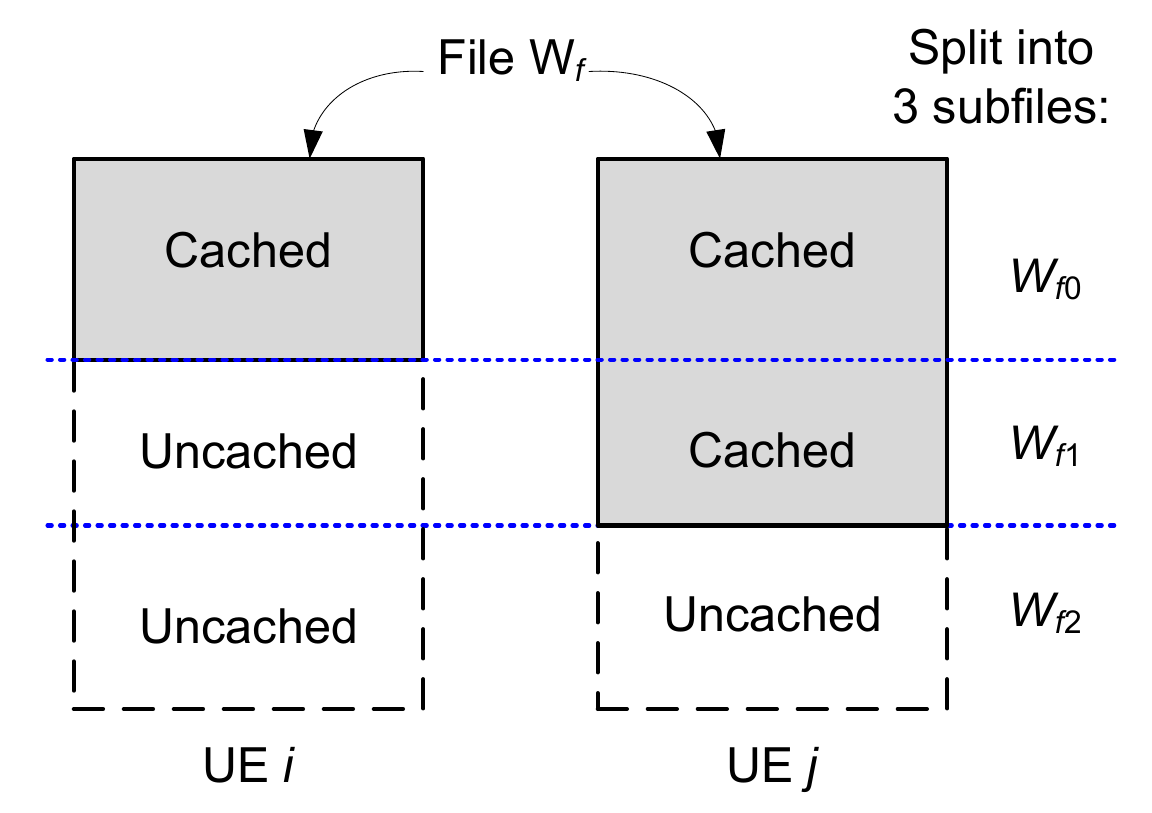}} 

\vspace{-.2cm}
\subfloat[]{\label{fig1b} \centering\includegraphics[height=2.4in]{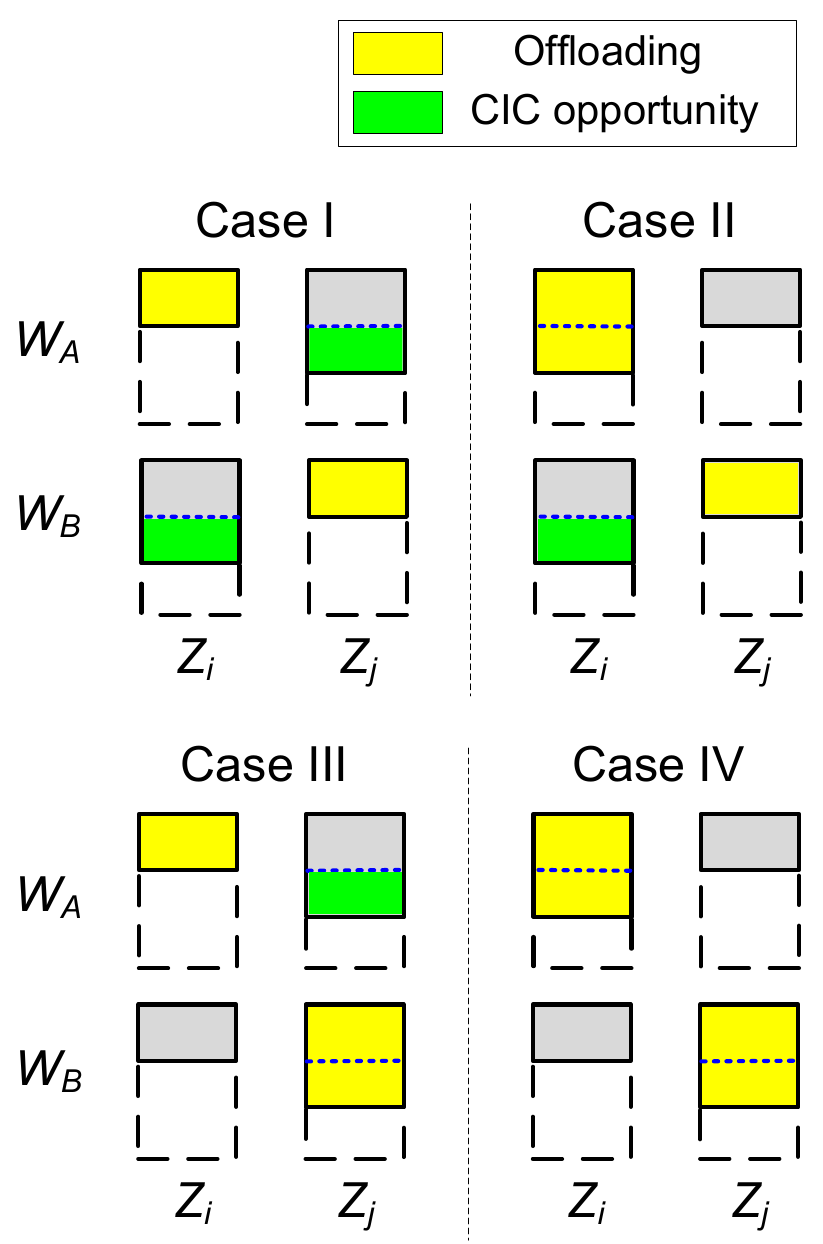}}
\caption{(a) Illustration of file splitting for cache-aided NOMA transmission: For prefix caching, the cached video chunks of file $W_f$, $f\in \{A,B\}$, at one UE are prefixes of those at the other UE, whereas, for general caching, each UE caches an arbitrary number of possibly different coded packets for file $W_f$. {{(b) Subfiles and cache status $Z_k$, $k\in\{i,j\}$, for different cache configurations. The portions of the files that can be offloaded and canceled via CIC are shown in yellow and green, respectively, cf. Remark~\ref{rem1}.}} }
\vspace{-.5cm} 
\end{figure}

Let $Z_{k}\triangleq(Z_{k,A},Z_{k,B})$, $k \in \{i,j\}$, denote the cache status of UE $k$, where $Z_{k,f}$, $f\in\left\{ A,B\right\}$, is the cached content of file $W_{f}$. We assume that $Z_{k,f}$, $f\in\left\{ A,B\right\}$, contains the first portion of the video chunks of $W_{f}$ in sequential order such that the initial playback delay is reduced if file $W_{f}$  is requested by UE $k$. Moreover, based on the users' requests and cache configurations, $W_{f}$ is split into three subfiles $(W_{f0},W_{f1},W_{f2})$ for adaptive file delivery. As illustrated in Fig.~\ref{fig1a}, $W_{f0}$ and $W_{f2}$ of size $\underline{c}_{f}V_{f}$ and $(1-\overline{c}_{f})V_{f}$ bits are the video chunks which are cached and uncached at both UEs, respectively, whereas subfile $W_{f1}$ of size $(\overline{c}_{f}-c_{kf})V_{f}$ bits is only cached at UE $\overline{k}_{f}$. Hence, we have $Z_{\underline{k}_{f},f}=(W_{f0})$ and $Z_{\overline{k}_{f},f}=(W_{f0},W_{f1})$, $f \in \{A,B\}$. The subfiles and cache status for the four possible cache configurations are illustrated in Fig.~\ref{fig1b}. As the cached data $Z_{\underline{k}_{f},f}$ is the prefix of $Z_{\overline{k}_{f},f}$, the considered caching scheme is referred to as \emph{prefix caching} in \cite{Choi:TWC16:Prefix}. 

\subsection{NOMA Transmission}

For video delivery, we assume a frequency flat quasi-static fading channel, where the channel coherence time exceeds the time needed for completion of file delivery. The received signal at UE $k \in \{i,j\}$ is given by 
\begin{equation}
y_{k}=h_{k}x+z_{k},
\end{equation}
where $h_{k}\in\mathbb{C}$ denotes the complex-valued channel gain between the BS and UE $k$, which is constant during the transmission of files $W_A$ and $W_B$. $x$ is the transmit signal and $z_{k}\sim\mathcal{CN}(0,\sigma_{k}^{2})$ is the AWGN at UE $k$.

The BS is assumed to know the cache statuses $Z_{i}$ and $Z_{j}$ during video delivery. Hence, the BS only transmits the uncached subfiles requested by the UEs. Thereby, four independent channel codebooks are employed to encode subfiles $W_{fs}$, $f\in \{A, B\}$, $s \in \{1,2\}$, at the BS and the corresponding codewords are superposed before being broadcast over the channel according to the NOMA principle. Taking the different cache configurations into account, the respective BS transmit signals are given by
\begin{equation}
x \!\!=\!\! 
\begin{cases}
\!\! \sqrt{p_{i,1}}x_{A1} \!+\! \sqrt{p_{i,2}}x_{A2} \!+\! \sqrt{p_{j,1}}x_{B1} \!+\! \sqrt{p_{j,2}}x_{B2}, \hfill \text{for Case \mbox{I}}, \\
\!\!\sqrt{p_{i,2}}x_{A2} + \sqrt{p_{j,1}}x_{B1}+\sqrt{p_{j,2}}x_{B2}, \hfill  \text{for Case \mbox{II}}, \\
\!\!\sqrt{p_{i,1}}x_{A1}+\sqrt{p_{i,2}}x_{A2} + \sqrt{p_{j,2}}x_{B2}, \hfill \text{for Case \mbox{III}}, \\
\!\!\sqrt{p_{i,2}}x_{A2} + \sqrt{p_{j,2}}x_{B2}, \hfill \text{for Case \mbox{IV}}, 
\end{cases}
\label{eq:txsig}
\end{equation}
where $x_{fs}$, $f\in\left\{ A,B\right\}$, $s\in \{1, 2\}$, is the codeword corresponding to subfile $W_{fs}$, and $\mathbb{E}\left[\left|x_{fs}\right|^{2}\right]=1$. $p_{k,s}\ge0$, $k \in \{i,j\}$, $s\in\left\{ 1,2\right\}$, denotes the transmit power of codeword $x_{fs}$. 
In \eqref{eq:txsig}, $x_{A1}$ ($x_{B1}$) is not transmitted in Cases \mbox{II} and \mbox{IV} (Cases \mbox{III} and \mbox{IV}) as it is already available at the requesting UE $i$ (UE $j$).

As the channel is static, we consider time-invariant power allocation, i.e., the powers, $p_{k,s}$, are fixed during file delivery. The total transmit power at the BS is constrained to $P$, i.e.,
\begin{equation}
\textrm{C1:}\;\sum\nolimits _{k\in\left\{ i,j\right\} }\sum\nolimits _{s\in\left\{ 1,2\right\} }p_{k,s}\le P.\label{eq:C1}
\end{equation}
For future reference, we define $\mathbf{p} \!\triangleq\! \left( p_{i,1},p_{i,2},p_{j,1},p_{j,2} \right)$ and $\mathcal{P} \!\triangleq\! \left\{ \mathbf{p}\in \mathbb{R}_{+}^{4} \mid \textrm{C1}  \right\}$ as the power allocation vector and the corresponding feasible set, respectively. 

\subsection{Joint CIC and SIC Decoding}

\begin{figure}[t]
\centering
\includegraphics[width=3.3in]{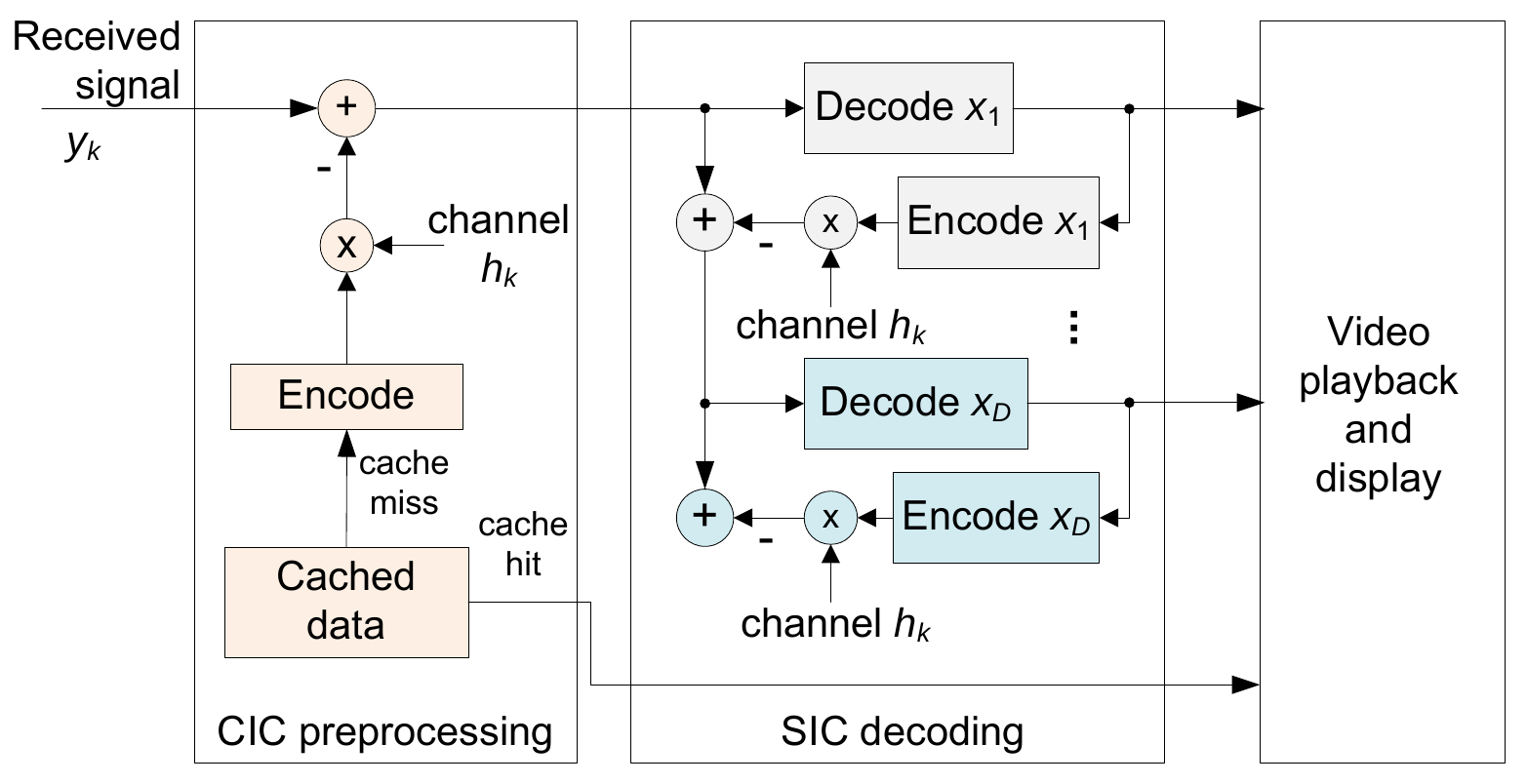}
\caption{\label{fig2a} Joint CIC and SIC decoding at UE $k\in \{i,j\} $ for cache-aided NOMA.  {{$x_1, \ldots, x_D$, $D \le 3$, represent the residual signals $x_{fs}$, $f\in \{A,B\}$, $s \in \{1,2\}$, which are not canceled by CIC but decoded successively by employing SIC. The order in which the $x_{fs}$s are decoded can be optimized.}}}
\vspace{-.5cm} 
\end{figure}

The proposed cache-aided NOMA scheme enables CIC at the receiver, which is not possible with conventional NOMA. The  joint CIC and SIC receiver performs CIC preprocessing of the received signal before SIC decoding as illustrated in Fig.~\ref{fig2a}. In particular, in Cases \mbox{I} and \mbox{III} (Cases \mbox{I} and \mbox{II}), the interference caused by codeword $x_{A1}$ ($x_{B1}$), which is requested by UE $i = \underline{k}_{A}$ ($j = \underline{k}_{B}$), is canceled at UE $j = \overline{k}_{A}$ ($i = \overline{k}_{B}$) by exploiting the cached data $Z_{\overline{k}_{A},A}$ ($Z_{\overline{k}_{B},B}$). Hence, the residual received signal after CIC preprocessing is given by
\begin{alignat}{1}
y_{i}^{\mathrm{CIC}} \!\!&=\!\! 
\begin{cases}
h_{i}(\sqrt{p_{i,1}}x_{A1}+\sqrt{p_{i,2}}x_{A2}+\sqrt{p_{j,2}}x_{B2})+z_{i}, \\
\hfill \text{for Cases \mbox{I} \& \mbox{III}}, \\
h_{i}(\sqrt{p_{i,2}}x_{A2}+\sqrt{p_{j,2}}x_{B2})+z_{i}, \\
\hfill \text{for Cases \mbox{II} \& \mbox{IV}}, 
\end{cases} 
\label{eq9}\\
y_{j}^{\mathrm{CIC}} \!\!&= \!\!
\begin{cases}
h_{j}(\sqrt{p_{i,2}}x_{A2} \!+\! \sqrt{p_{j,1}}x_{B1} \!+\! \sqrt{p_{j,2}}x_{B2})+z_{j}, \\
 \hfill \text{for Cases \mbox{I} \& \mbox{II}}, \\
h_{j}(\sqrt{p_{i,2}}x_{A2}+ \sqrt{p_{j,2}}x_{B2})+z_{j}, \\
 \hfill \text{for Cases \mbox{III} \& \mbox{IV}}. 
\end{cases}
\label{eq10} 
\end{alignat}

\begin{rem} \label{rem1}
{{As shown in Fig.~\ref{fig1b}, for the proposed cache-aided NOMA scheme, $c_{kf}$ portion of file $W_f$ is not transmitted at all as it is already cached at the requesting UE, while $\overline{c}_{f}-c_{kf}$ portion of file $W_f$ can be removed from the received signal of the non-requesting UE $k'$ via CIC, where $(k,f)\in\left\{ (i,A),(j,B)\right\}$ and $k'\neq k$.}} As such, the proposed scheme can exploit $\overline{c}_f$ portion of file $W_f$ for performance improvement, even if the unfavorable cache configuration of Case \mbox{I} occurs, whereas a straightforward combination of caching and NOMA can only exploit the cached portion $c_{kf}$  of the requested file \cite{Ding17Noma:Caching}. 
\end{rem}

\begin{rem}\label{rem2}
As CIC reduces multiuser interference, multiple decoding orders may become feasible for SIC processing of $y_{k}^{\mathrm{CIC}}$, $k \in \{i, j\}$. For example, for Case \mbox{I}, there are $4!=24$ candidate decoding orders based on \eqref{eq9} and \eqref{eq10} compared to only $2!=2$ candidate decoding orders (of which only one is feasible) for conventional NOMA. This leads to a substantially increased flexibility for decoding the video data based on $y_{k}^{\mathrm{CIC}}$. Hence, different from conventional downlink NOMA transmission, for cache-aided NOMA, joint optimization of the SIC decoding order and the power and rate allocation based on the cache status and channel conditions is desirable for performance optimization. Although decoding order optimization is in principle a combinatorial problem, we show in Section~\ref{sec3} by careful inspection of the SIC decoding condition that the optimal decoding order is contained in a small subset of all possible decoding orders, and hence the associated complexity is limited. 
\end{rem}

\subsection{\label{sec2-4}Extensions}
\subsubsection{Extension to General Caching Schemes}
The proposed cache-aided NOMA scheme is also applicable to general caching methods other than prefix caching. In particular, by employing additional intra-session source coding of the video files using e.g. maximum distance separable (MDS) codes \cite{Caire13IT:FemtoCaching,Land:Globecom15:MDS,KKWong:TWC17}, the UEs can perform cache-aided NOMA even if arbitrary rather than the first portions of the video chunks of the video files are cached, which increases the flexibility in the caching phase. Moreover, different from wireless coded caching \cite{Timo:ISWCS15:Joint,Shirin17CacheAssign,Petros:ISIT17,Tulino:JSAC16:Channel-aware}, MDS source coding is independent of the channel coding employed for NOMA transmission and can be efficiently implemented in practical systems.

Assume that an $\left(m_{P},\,m_{C}\right)$ MDS source code is employed for file $W_f$, $f\in \{A,B\}$, where $m_{P}\in\mathbb{N}$ packets are created from $m_{C}\in\mathbb{N}$ chunks of the video file. The coded packets are independent. Moreover, any $m_{C}(1+\epsilon)$ unique coded packets collected from the cached and received data are sufficient to recover the original video data for $\epsilon\ge0$. Let $m_{if}$ and $m_{jf}$, $f \in \{A, B\}$, be the number of (arbitrary) packets of $W_f$ that have already been cached at UEs $i$ and $j$, respectively. Without loss of generality, assume $m_{if} < m_{jf}$. Then, a minimum number of $m_{if}$ possibly different packets are cached at both UEs and can be offloaded once they are requested. Let $W_{f0}$ denote the packets offloaded by caching. Different from prefix caching, now the packets of $W_{f0}$ may vary across the UEs. Moreover, for file $W_A$, at least $m_{jA}-m_{iA}$ unique packets are cached at UE $j$, which are not cached at UE $i$ but have to be newly delivered as subfile $W_{A1}$. For file $W_B$, subfile $W_{B1}$ is defined in the same manner. Finally, for successful decoding of the video files, the remaining $m_{C}(1+\epsilon)-m_{jf}$ packets, which are contained in subfiles $W_{A2}$ and $W_{B2}$, also have to be delivered to the requesting UE.

Hence, subfiles $W_{fs}$, $f\in \{A, B\}$, $s \in \{1,2\}$, are uniquely defined for both UEs and the definition of $Z_i$ and $Z_j$ based on $W_{fs}$ remains the same as for prefix caching except that now the $W_{fs}$ consist of coded packets. By employing the proposed cache-aided NOMA scheme, subfile $W_{f1}$, $f\in \{A, B\}$, which is cached at the non-requesting UE, can be used for CIC. Note that, for (asymptotically) optimal MDS codes, the performance gap between general caching and prefix caching is negligible as $\epsilon\to 0$. Henceforth, in the remainder of this paper, we focus on analyzing cache-aided NOMA without specifying the caching scheme adopted.

\subsubsection{Extension to Multiple Users}
As commonly assumed in the NOMA literature \cite{Ding16UserP:TVT,Sun16FD-MC-NOMA:TCOM,Ding16MIMO-NOMA:TWC}, we  consider a system of two UEs to limit the decoding cost and delay incurred at the UE performing SIC. If more than two UEs are present, they can be grouped such that each group includes two UEs which share the same resource block via NOMA, while UEs in different groups are allocated to different resource blocks \cite{Ding16UserP:TVT,Sun16FD-MC-NOMA:TCOM,Ding16MIMO-NOMA:TWC}. Consequently, the proposed cache-aided NOMA scheme and the derived results can be extended to the case of more than two UEs in a straightforward manner. On the other hand, pairing more than two UEs in one group, although possible, may be undesirable because of the associated increase in complexity and delay. Furthermore, in practice, SIC decoders are imperfect and the resulting decoding errors  may propagate over multiple decoding stages, leading to additional performance degradation.

\section{\label{sec3}Achievable Rate Region}
 
In this section, we evaluate the achievable rate region of the proposed cache-aided NOMA scheme\footnote{{We note that the achievable rate region of downlink broadcast channels for the case where one user's message is \emph{fully} known at the other user(s) has been studied in \cite{Ong:IT15}. However, for cache-aided NOMA, a user's message may only be \emph{partially} known at the other user, cf. \eqref{eq9} and \eqref{eq10}. Due to this difference, the results in \cite{Ong:IT15} are not directly applicable to cache-aided NOMA.}}.
Based on the derived achievable rate region results, we further show that the Pareto optimal boundary can be determined by solving a rate maximization problem. Let $\mathbf{r}\triangleq \left( r_{i,1},r_{i,2},r_{j,1},r_{j,2} \right)$ be the rate allocation vector, where  $r_{k,s}\ge0$ is the rate for delivering $x_{fs}$ to UE $k$, $(k,f) \in \{(i,A),(j,B)\}$, $s\in\left\{ 1,2\right\}$. We define $\alpha_{k}\triangleq\frac{\sigma_{k}^{2}}{\left|h_{k}\right|^{2}}$, $k \in \{i, j\}$, as the effective noise variance at UE $k$. Without loss of generality, we assume $\alpha_{i}<\alpha_{j}$, i.e., UE $i$ has a larger channel gain than UE $j$. 

\subsection{\label{sub3-1}Derivation of Achievable Rate Region}
We first consider Case \mbox{I}, which among the four possible cache configurations is the most challenging to analyze. We then show that the results obtained for Case \mbox{I} can be extended to Cases \mbox{II}--\mbox{IV} in a straightforward manner.

According to \eqref{eq9} and \eqref{eq10}, for Case \mbox{I}, signals $x_{f1}$ and $x_{f2}$ are delivered to UE $k$ simultaneously, where $(k,f) \in \{(i,A), (j,B)\}$,  and $x_{B1}$ ($x_{A1}$) is canceled at UE $i$ (UE $j$) by CIC. Moreover, $x_{A2}$ and $x_{B2}$, which constitute interference signals at one of the UEs, are received at both UEs, whereas $x_{A1}$ and $x_{B1}$ are only included in the residual received signal $y_{k}^{\mathrm{CIC}}$ of the requesting UE. The interference signals can be decoded and canceled only if the SIC decoding condition is fulfilled, i.e., the received SINR for $x_{A2}$ and $x_{B2}$ at the non-requesting UEs, UE $j$ and UE $i$, has to exceed that at the requesting UEs, UE $i$ and UE $j$, respectively. In contrast, for signals $x_{A1}$ and $x_{B1}$, such a constraint is not required. Depending on which signal is decoded first, three cases can be distinguished: for the first two cases, signals $x_{A1}$ and $x_{B1}$ are decoded first at the requesting UEs, respectively, whereas, for the third case, the interference signals $x_{A2}$ and $x_{B2}$ are decoded first. Since \eqref{eq9} and \eqref{eq10} constitute a non-degraded broadcast channel \cite{Cover12IT-Elements,Gamal11NIT}, the achievable rate regions for all three cases have to be evaluated for specific power regions individually. 

\subsubsection{\label{enu:IV-1}UE $i\overset{_{(1)}}{\to}x_{A1}$}
 If UE $i$ decodes $x_{A1}$ first, signals $y_{i}^{\mathrm{_{(1)}}}=h_{i}(\sqrt{p_{i,2}}x_{A2}+\sqrt{p_{j,2}}x_{B2})+z_{i}$ and $y_{j}^{\mathrm{_{(1)}}}=h_{j}(\sqrt{p_{i,2}}x_{A2}+\sqrt{p_{j,2}}x_{B2}+\sqrt{p_{j,1}}x_{B1})+z_{j}$ have to be subsequently decoded at UE $i$ and UE $j$, respectively. The achievable rate region is provided in Proposition \ref{prop1}.
\vspace{-.2cm}
\begin{prop}
\emph{\label{prop1} When UE $i\overset{_{(1)}}{\to}x_{A1}$, the rate region $ \mathcal{R}_{\lyxmathsym{\mbox{I}},1}\left(\mathcal{P}_{\lyxmathsym{\mbox{I}},1}\right)\bigcup\mathcal{R}_{\lyxmathsym{\mbox{I}},2}\left(\mathcal{P}_{\lyxmathsym{\mbox{I}},2}\right)$ is achievable, where 
\vspace{-.2cm}
\begin{alignat}{1}
\mathcal{R}_{\lyxmathsym{\mbox{I}},1}\left(\mathcal{P}_{\lyxmathsym{\mbox{I}},1}\right) & \!\! \triangleq \!\!\!\! \bigcup_{\mathbf{p}\in\mathcal{P}_{\lyxmathsym{\mbox{I}},1}} \!\! \left\{\! \mathbf{r}\left| \!\! \begin{array}{l}
r_{i,1}\le C_{i,1}^{_{[1]}} = C\left(\tfrac{p_{i,1}}{p_{i,2}+p_{j,2}+\alpha_{i}}\right), \\  r_{i,2}\le C_{i,2}^{_{[1]}} = C\left(\tfrac{p_{i,2}}{\alpha_{i}}\right), \\
r_{j,s}\le C_{j,s}^{_{[1]}} = C\left(\tfrac{p_{j,s}}{p_{i,2}+\alpha_{j}}\right), s \!\in\! \left\{1,2\right\},\\ 
r_{j,1}+r_{j,2}\le C_{j,1,2}^{_{[1]}} = C\left(\tfrac{p_{j,1}+p_{j,2}}{p_{i,2}+\alpha_{j}}\right), 
\end{array} \!\!\!\!\! \right.\right\} \label{eq:rate-II-1-1}\\
\mathcal{R}_{\lyxmathsym{\mbox{I}},2}\left(\mathcal{P}_{\lyxmathsym{\mbox{I}},2}\right) & \!\!\triangleq \!\!\! \bigcup_{\mathbf{p}\in\mathcal{P}_{\lyxmathsym{\mbox{I}},2}} \!\! \left\{ \!\mathbf{r}\left| \!\! \begin{array}{l}
r_{i,1}\le C_{i,1}^{_{[2]}} = C\left(\tfrac{p_{i,1}}{p_{i,2}+p_{j,2}+\alpha_{i}}\right),\\ r_{i,2}\le C_{i,2}^{_{[2]}} = C\left(\tfrac{p_{i,2}}{p_{j,2}+\alpha_{i}}\right),\\
r_{j,1}\le C_{j,1}^{_{[2]}} = C\left(\tfrac{p_{j,1}}{\alpha_{j}}\right),\\ r_{j,2}\le C_{j,2}^{_{[2]}} = C\left(\tfrac{p_{j,2}}{p_{i,2}+p_{j,1}+\alpha_{j}}\right),
\end{array} \!\!\!\!\! \right.\right\} \label{eq:rate-II-2-1}
\end{alignat}
with $\mathcal{P}_{\lyxmathsym{\mbox{I}},1} = \mathcal{P}$ and $\mathcal{P}_{\lyxmathsym{\mbox{I}},2}\triangleq\left\{ \mathbf{p}\in\mathcal{P}\mid p_{j,2}-p_{j,1}>\alpha_{j}-\alpha_{i}\right\} $. For $\mathcal{R}_{\lyxmathsym{\mbox{I}},1}\left(\mathcal{P}_{\lyxmathsym{\mbox{I}},1}\right)$, the decoding orders for UEs $i$ and $j$ are given as $i\overset{_{(1)}}{\to}x_{A1}\overset{_{(2)}}{\to}x_{B2}\overset{_{(3)}}{\to}x_{A2}$ and $j\overset{_{(1)}}{\to}\left(x_{B1},\,x_{B2}\right)$, respectively.  For $\mathcal{R}_{\lyxmathsym{\mbox{I}},2}\left(\mathcal{P}_{\lyxmathsym{\mbox{I}},2}\right)$, the decoding orders are $i\overset{_{(1)}}{\to}x_{A1}\overset{_{(2)}}{\to}x_{A2}$ and $j\overset{_{(1)}}{\to}x_{A2}\overset{_{(2)}}{\to}x_{B2}\overset{_{(3)}}{\to}x_{B1}$, respectively. }
\end{prop} 
\begin{IEEEproof}
Please refer to the Appendix.
\end{IEEEproof}

\begin{rem}
In Proposition~\ref{prop1}, the interference for decoding $x_{A2}$ is reduced after $x_{A1}$ has been decoded and canceled from $y_{i}^{\mathrm{CIC}}$. Hence, decoding $x_{A1}$ first is desirable when e.g. $W_{A1}$ has a smaller size and/or requires a lower delivery rate than $W_{A2}$. Moreover, UE $j$ can achieve a larger rate in $\mathcal{R}_{\lyxmathsym{\mbox{I}},2}\left(\mathcal{P}_{\lyxmathsym{\mbox{I}},2}\right)$ than in $\mathcal{R}_{\lyxmathsym{\mbox{I}},1}\left(\mathcal{P}_{\lyxmathsym{\mbox{I}},1}\right)$ as 
\begin{equation}
C_{j,1}^{_{[2]}} + C_{j,2}^{_{[2]}} \!\ge\! C \!\left(\frac{p_{j,1}}{p_{i,2} + \alpha_{j}}\right) + C \!\left(\frac{p_{j,2}}{p_{i,2}+p_{j,1}+\alpha_{j}}\right) \!=\! C_{j,1,2}^{_{[1]}},
\label{eq:ratecomp}
\end{equation}
where equality holds for $p_{i,2} =0$.
\end{rem}

\subsubsection{\label{enu:IV-2}UE $j\overset{_{(1)}}{\to}x_{B1}$}

Decoding and canceling $x_{B1}$ first improves the SINR of $x_{B2}$ at UE $j$, which is desirable when subfile $W_{B1}$ has a smaller size than $W_{B2}$. After $x_{B1}$ has been canceled, the resulting signals at UE $i$ and UE $j$ are $y_{i}^{\mathrm{_{(1)}}}=h_{i}(\sqrt{p_{i,1}}x_{A1}+\sqrt{p_{i,2}}x_{A2}+\sqrt{p_{j,2}}x_{B2})+z_{i}$ and $y_{j}^{\mathrm{_{(1)}}}=h_{j}(\sqrt{p_{i,2}}x_{A2}+\sqrt{p_{j,2}}x_{B2})+z_{j}$, respectively. The corresponding achievable rate region is given in Proposition~\ref{prop2}.

\vspace{-0.2cm}
\begin{prop}
\emph{\label{prop2} When UE $j\overset{_{(1)}}{\to}x_{B1}$, the  rate region $\mathcal{R}_{\text{\mbox{I}},3}\left(\mathcal{P}_{\lyxmathsym{\mbox{I}},3}\right) \bigcup \mathcal{R}_{\text{\mbox{I}},4}\left(\mathcal{P}_{\lyxmathsym{\mbox{I}},4}\right) \bigcup \mathcal{R}_{\text{\mbox{I}},5}\left(\mathcal{P}_{\lyxmathsym{\mbox{I}},5}\right)$ is achievable, where 
\begin{alignat}{1}
 & \mathcal{R}_{\text{\mbox{I}},3}\left(\mathcal{P}_{\lyxmathsym{\mbox{I}},3}\right)\triangleq \!\! \bigcup_{\mathbf{p}\in\mathcal{P}_{\lyxmathsym{\mbox{I}},3}} \!\! \left\{ \mathbf{r}\left|\begin{array}{l}
r_{i,s}\le C\left(\frac{p_{i,s}}{\alpha_{i}}\right),\,s\in\left\{ 1,2\right\} ,\\ r_{i,1}+r_{i,2}\le C\left(\frac{p_{i,1}+p_{i,2}}{\alpha_{i}}\right),\\
r_{j,1}\le C\left(\frac{p_{j,1}}{p_{i,2}+p_{j,2}+\alpha_{j}}\right),\\ r_{j,2}\le C\left(\frac{p_{j,2}}{p_{i,2}+\alpha_{j}}\right),
\end{array} \!\!\!
\right.\right\}  \label{eq:rate-III-1} \\
 & \mathcal{R}_{\text{\mbox{I}},4}\left(\mathcal{P}_{\lyxmathsym{\mbox{I}},4}\right)\triangleq\bigcup_{\mathbf{p}\in \mathcal{P}_{\lyxmathsym{\mbox{I}},4}}\left\{ \mathbf{r}\left|\begin{array}{l}
r_{i,1}\le C\left(\frac{p_{i,1}}{p_{j,2}+\alpha_{i}}\right),\\ r_{i,2}\le C\left(\frac{p_{i,2}}{p_{i,1}+p_{j,2}+\alpha_{i}}\right),\\
r_{j,1}\le C\left(\frac{p_{j,1}}{p_{i,2}+p_{j,2}+\alpha_{j}}\right),\\ r_{j,2}\le C\left(\frac{p_{j,2}}{\alpha_{j}}\right),\\
\end{array}\right.\right\} \label{eq:rate-III-2} \\
 & \mathcal{R}_{\text{\mbox{I}},5}\left(\mathcal{P}_{\lyxmathsym{\mbox{I}},5}\right)\triangleq\bigcup_{\mathbf{p}\in \mathcal{P}_{\lyxmathsym{\mbox{I}},5}}\left\{ \mathbf{r}\left|\begin{array}{l}
r_{i,1}\le C\left(\frac{p_{i,1}}{p_{i,2}+p_{j,2}+\alpha_{i}}\right),\\ r_{i,2}\le C\left(\frac{p_{i,2}}{\alpha_{i}}\right),\\
r_{j,1}\le C\left(\frac{p_{j,1}}{p_{i,2}+p_{j,2}+\alpha_{j}}\right),\\ r_{j,2}\le C\left(\frac{p_{j,2}}{p_{i,2}+\alpha_{j}}\right),
\end{array}\right.\right\} \label{eq:rate-III-3}
\end{alignat}
with $\mathcal{P}_{\lyxmathsym{\mbox{I}},3}\triangleq\left\{ \mathbf{p}\in\mathcal{P}\mid p_{i,1}<\alpha_{j}-\alpha_{i}\right\}$ and  $\mathcal{P}_{\lyxmathsym{\mbox{I}},4} = \mathcal{P}_{\lyxmathsym{\mbox{I}},5} = \mathcal{P} \backslash \mathcal{P}_{\lyxmathsym{\mbox{I}},3}$. The decoding orders achieving $\mathcal{R}_{\text{\mbox{I}},3}\left(\mathcal{P}_{\lyxmathsym{\mbox{I}},3}\right)$ are UE $i\overset{_{(1)}}{\to}x_{B2}\overset{_{(2)}}{\to}(x_{A1},x_{A2})$ and UE $j\overset{_{(1)}}{\to}x_{B1}\overset{_{(2)}}{\to}x_{B2}$.  Moreover, the decoding orders for $\mathcal{R}_{\text{\mbox{I}},4}\left(\mathcal{P}_{\lyxmathsym{\mbox{I}},4}\right)$ are UE $i\overset{_{(1)}}{\to}x_{A2}\overset{_{(2)}}{\to}x_{A1}$ and UE $j\overset{_{(1)}}{\to}x_{B1}\overset{_{(2)}}{\to}x_{A2}\overset{_{(3)}}{\to}x_{B2}$. Finally, $\mathcal{R}_{\text{\mbox{I}},5}\left(\mathcal{P}_{\lyxmathsym{\mbox{I}},5}\right)$ is achieved by the decoding orders UE $i\overset{_{(1)}}{\to}x_{A1}\overset{_{(2)}}{\to}x_{B2}\overset{_{(3)}}{\to}x_{A2}$ and UE $j\overset{_{(1)}}{\to}x_{B1}\overset{_{(2)}}{\to}x_{B2}$. }
\end{prop}
\begin{IEEEproof}
The proof is similar to that for Proposition~\ref{prop1}.
\end{IEEEproof}
\begin{rem}
From Propositions~\ref{prop1} and~\ref{prop2} we have $\mathcal{R}_{\text{\mbox{I}},5}\left(\mathcal{P}_{\lyxmathsym{\mbox{I}},5}\right)\subseteq\mathcal{R}_{\text{\mbox{I}},1}\left(\mathcal{P}_{\text{\mbox{I}},1}\right)$, where the decoding orders for achieving both rate regions coincide. Hence, for Case \mbox{I}, $\mathcal{R}_{\text{\mbox{I}},5}\left(\mathcal{P}_{\lyxmathsym{\mbox{I}},5}\right)$ can be ignored without affecting the overall achievable rate region. However, $\mathcal{R}_{\text{\mbox{I}},5}\left(\mathcal{P}_{\lyxmathsym{\mbox{I}},5}\right)$ is needed to obtain the achievable rate region for Case \mbox{III}, cf. $\mathcal{R}_{\text{\mbox{III}}}(\mathcal{P})$ in Corollary~\ref{cor1}, and hence, is included here. 
\end{rem}

\subsubsection{\label{enu:IV-3}UE $j\overset{_{(1)}}{\to}(x_{A2},x_{B2})$ and UE $i\overset{_{(1)}}{\to}(x_{A2},x_{B2})$}
Recall that decoding the interference signals first is only possible if the SIC condition is fulfilled. For this case, the achievable rate region is given in Proposition \ref{prop3}.

\begin{prop}
\emph{\label{prop3} When UE $j\overset{_{(1)}}{\to}(x_{A2},x_{B2})$ and UE $i\overset{_{(1)}}{\to}(x_{A2},x_{B2})$, the achievable rate region is given by $\mathcal{R}_{\text{\mbox{I}},6}\left(\mathcal{P}_{\lyxmathsym{\mbox{I}},6}\right) \bigcup \mathcal{R}_{\text{\mbox{I}},7}\left(\mathcal{P}_{\lyxmathsym{\mbox{I}},7}\right) \bigcup \mathcal{R}_{\text{\mbox{I}},8}\left(\mathcal{P}_{\lyxmathsym{\mbox{I}},8}\right),$ where
\begin{alignat}{1}
 & \mathcal{R}_{\text{\mbox{I}},6}\left(\mathcal{P}_{\lyxmathsym{\mbox{I}},6}\right)\triangleq \!\! \bigcup_{\mathbf{p}\in\mathcal{P}_{\lyxmathsym{\mbox{I}},6}} \!\! \left\{ \mathbf{r}\left|\begin{array}{l}
r_{i,s}\le C\left(\frac{p_{i,s}}{\alpha_{i}}\right),\,s\in\left\{ 1,2\right\} ,\\
r_{i,1}+r_{i,2}\le C\left(\frac{p_{i,1}+p_{i,2}}{\alpha_{i}}\right),\\
r_{j,1}\le C\left(\frac{p_{j,1}}{p_{i,2}+\alpha_{j}}\right),\\ 
r_{j,2}\le C\left(\frac{p_{j,2}}{p_{j,1}+p_{i,2}+\alpha_{j}}\right).
\end{array} \!\!\!\! \right.\right\} \label{eq:rate-IV-3-1}\\
 & \mathcal{R}_{\text{\mbox{I}},7}\left(\mathcal{P}_{\lyxmathsym{\mbox{I}},7}\right)\triangleq \!\! \bigcup_{\mathbf{p}\in \mathcal{P}_{\lyxmathsym{\mbox{I}},7}}\left\{ \mathbf{r}\left|\begin{array}{l}
r_{i,1}\le C\left(\frac{p_{i,1}}{\alpha_{i}+p_{j,2} \Delta }\right), \\
r_{i,2}\le C\left(\frac{p_{i,2}}{p_{i,1}+p_{j,2}+\alpha_{i}}\right), \\
r_{j,1}\le C\left(\frac{p_{j,1}}{\alpha_{j}}\right),\\
r_{j,2}\le C\left(\frac{p_{j,2}}{p_{i,2}+p_{j,1}+\alpha_{j}}\right),
\end{array} \!\! \right.\right\} \label{eq:rate-IV-3-2} \\
 & \mathcal{R}_{\text{\mbox{I}},8}\left(\mathcal{P}_{\lyxmathsym{\mbox{I}},8}\right)\triangleq \!\! \bigcup_{\mathbf{p}\in \mathcal{P}_{\lyxmathsym{\mbox{I}},8}} \!\! \left\{ \mathbf{r}\left|\begin{array}{l}
r_{i,1}\le C\left(\frac{p_{i,1}}{p_{j,2}+\alpha_{i}}\right),\\ 
r_{i,2}\le C\left(\frac{p_{i,2}}{p_{i,1}+p_{j,2}+\alpha_{i}}\right),\\
r_{j,s}\le C\left(\frac{p_{j,s}}{\alpha_{j}}\right), s\in\left\{ 1,2\right\} ,\\
 r_{j,1}+r_{j,2}\le C\left(\frac{p_{j,1}+p_{j,2}}{\alpha_{j}}\right),
\end{array} \!\!\!\! \right.\right\} \label{eq:rate-IV-3-3}
\end{alignat}
with $\mathcal{P}_{\lyxmathsym{\mbox{I}},6}\triangleq\left\{ \mathbf{p}\in\mathcal{P}\mid p_{i,1}<p_{j,1}+\alpha_{j}-\alpha_{i}\right\}$, $\mathcal{P}_{\lyxmathsym{\mbox{I}},7} = \mathcal{P}_{\lyxmathsym{\mbox{I}},8} = \mathcal{P} \backslash \mathcal{P}_{\lyxmathsym{\mbox{I}},6}$, and $\Delta \triangleq \mathbf{1} [p_{i,2} > p_{i,1}-p_{j,1}-\alpha_{j}+\alpha_{i}]$. The decoding orders achieving $\mathcal{R}_{\text{\mbox{I}},6}\left(\mathcal{P}_{\lyxmathsym{\mbox{I}},6}\right)$ are UE $i\overset{_{(1)}}{\to}x_{B2}\overset{_{(2)}}{\to}(x_{A1},x_{A2})$ and UE $j\overset{_{(1)}}{\to}x_{B2}\overset{_{(2)}}{\to}x_{B1}$. Moreover, $\mathcal{R}_{\text{\mbox{I}},7}\left(\mathcal{P}_{\lyxmathsym{\mbox{I}},7}\right)$ is achieved with decoding orders
\[
\textrm{UE}\;i\overset{_{(1)}}{\to}x_{A2}\overset{_{(2)}}{\to}\begin{cases}
x_{A1}, & \textrm{if } \Delta = 1,\\
x_{B2}\overset{_{(3)}}{\to}x_{A1}, & \textrm{otherwise,}
\end{cases}
\]
 and UE $j\overset{_{(1)}}{\to}x_{B2}\overset{_{(2)}}{\to}x_{B1}$. Finally, $\mathcal{R}_{\text{\mbox{I}},8}\left(\mathcal{P}_{\lyxmathsym{\mbox{I}},8}\right)$ is achieved with decoding orders UE $i\overset{_{(1)}}{\to}x_{A2}\overset{_{(2)}}{\to}x_{A1}$ and UE $j\overset{_{(1)}}{\to} x_{A2}\overset{_{(2)}}{\to}(x_{B1},x_{B2})$.}
\end{prop}
\begin{IEEEproof}
The proof is similar to that for Proposition~\ref{prop1}.
\end{IEEEproof}

\begin{rem}
Different from conventional NOMA, for the proposed cache-aided NOMA scheme, file splitting enables joint decoding. For example, joint decoding of $x_{B1}$ and $x_{B2}$ at UE $j$ is possible in $\mathcal{R}_{\lyxmathsym{\mbox{I}}, 1}\left(\mathcal{P}_{\lyxmathsym{\mbox{I}}, 1}\right)$ and $\mathcal{R}_{\lyxmathsym{\mbox{I}}, 8}\left(\mathcal{P}_{\lyxmathsym{\mbox{I}}, 8}\right)$, as the two signals are received by UE $j$ over the same AWGN channel with noise variances $p_{i,2} + \alpha_j$ and $\alpha_j$, respectively. Therefore, UE $j$ can flexibly choose the decoding order of these files. Similarly, joint decoding of $x_{A1}$ and $x_{A2}$ is possible at UE $i$ in $\mathcal{R}_{\lyxmathsym{\mbox{I}}, 3}\left(\mathcal{P}_{\lyxmathsym{\mbox{I}}, 3}\right)$ and $\mathcal{R}_{\lyxmathsym{\mbox{I}}, 6}\left(\mathcal{P}_{\lyxmathsym{\mbox{I}}, 6}\right)$, respectively. We note that employing file splitting in conventional NOMA cannot increase the achievable rates at the UEs. However, for the proposed cache-aided NOMA, if a portion of a file is cached at one of the UEs, the achievable rates of the UEs can be increased by file splitting because of the CIC.   
\end{rem}

Finally, combining the results in Propositions~\ref{prop1}--\ref{prop3}, the overall achievable rate region for Case \mbox{I} is given by $\mathcal{R}_{\text{\mbox{I}}} (\mathcal{P}) \triangleq \bigcup_{n=1}^{8} \mathcal{R}_{\text{\mbox{I}},n} (\mathcal{P}_{\text{\mbox{I}},n})$.  Furthermore, the achievable rate regions for the other cache configurations can be obtained as special cases of $\mathcal{R}_{\text{\mbox{I}}}$, as shown in Corollary \ref{cor1}.  

\begin{cor}
\emph{For Cases \mbox{II}, \mbox{III}, and \mbox{IV}, the respective achievable rate regions are given by
\begin{alignat}{1}
\mathcal{R}_{\text{\mbox{II}}}(\mathcal{P}) &=\bigcup_{\mathbf{p}\in\mathcal{P}}\left\{ \mathbf{r}\mid\mathbf{r}\in \bigcup\nolimits_{n=1}^{2} \mathcal{R}_{\text{\mbox{I}},n} (\mathcal{P}_{\text{\mbox{I}},n}), \,r_{i,1}=0\right\}, \label{eq24} \\
\mathcal{R}_{\text{\mbox{III}}}(\mathcal{P}) &=\bigcup_{\mathbf{p}\in\mathcal{P}}\left\{ \mathbf{r}\mid\mathbf{r}\in \bigcup\nolimits_{n=3}^{5} \mathcal{R}_{\text{\mbox{I}},n} (\mathcal{P}_{\text{\mbox{I}},n}),\,r_{j,1}=0\right\}, \label{eq25} \\
\mathcal{R}_{\text{\mbox{IV}}}(\mathcal{P}) &=\bigcup_{\mathbf{p}\in\mathcal{P}}\left\{ \left(r_{i,2},r_{j,2}\right) \left|\begin{array}{l}
r_{i,2}\le C\left(\frac{p_{i,2}}{\alpha_{i}}\right),\\
r_{j,2}\le C\left(\frac{p_{j,2}}{p_{i,2}+\alpha_{j}}\right)\end{array} \!\!\! \right.\right\} ,\label{eq26}   
\end{alignat} 
where \eqref{eq26} is a special case of \eqref{eq24} and \eqref{eq25} as $\mathcal{R}_{\text{\mbox{IV}}}(\mathcal{P}) = \mathcal{R}_{\text{\mbox{II}}}(\mathcal{P}) \bigcap \mathcal{R}_{\text{\mbox{III}}}(\mathcal{P})$.}
\label{cor1}
\end{cor}
\begin{IEEEproof}
From \eqref{eq9} and \eqref{eq10} we observe that the received signals after CIC for Case \mbox{II} (Case \mbox{III}) are identical to those for Case \mbox{I} with UE $i\overset{_{(1)}}{\to}x_{A1}$ (UE $j\overset{_{(1)}}{\to}x_{B1}$). Hence, according to Propositions \ref{prop1} and \ref{prop2}, the achievable rate regions for Cases \mbox{II} and \mbox{III} are given by \eqref{eq24} and \eqref{eq25}, respectively, where $r_{i,1} = 0$ and $r_{j,1} = 0$, i.e., $p_{i,1} = 0$ and $p_{j,1} = 0$ are optimal. Finally, Case \mbox{IV} corresponds to a \emph{degraded} broadcast channel, and its capacity region \eqref{eq26} is achieved by SIC where $x_{B2}$ is canceled at UE $i$ before decoding $x_{A2}$, as $\alpha_{i}<\alpha_{j}$,  while $x_{A2}$ is treated as noise at UE $j$ for decoding $x_{B2}$ \cite{Cover12IT-Elements,Gamal11NIT}. This completes the proof. 
\end{IEEEproof}

\subsection{\label{sub3-3}Pareto Optimal Rate Tuples}

For a unified analysis, in the following, we drop the cache configuration index and denote the achievable rate region of cache-aided NOMA as $\mathcal{R} (\mathcal{P} ) =\bigcup_{n \in \mathcal{N}} \mathcal{R}_{n}(\mathcal{P}_{n})$, where $\mathcal{R}_{n}(\mathcal{P}_{n})$ is the rate region achieved by decoding order $n \in \mathcal{N} \subseteq \{1,\ldots,8\}$ as specified in Section~\ref{sub3-1}. To unify our presentation, we express $\mathcal{R}_{n}(\mathcal{P}_{n})$ as
\begin{equation}
\mathcal{R}_{n}(\mathcal{P}_{n}) \!= \!\! \bigcup_{\mathbf{p}\in\mathcal{P}_{n}} \!\! \left\{ \mathbf{r}\left| \!\!
\begin{array}{l}
\textrm{C2:}\,r_{ks} \!\le\! C_{k,s}^{_{[n]}}(\mathbf{p}), k \!\in\! \left\{ i,j\right\}, s \!\in\! \left\{ 1,2\right\} \\
\textrm{C3:}\,r_{k1} \!+\! r_{k2} \!\le\! C_{k,1,2}^{_{[n]}}(\mathbf{p}), k \!\in\! \left\{ i,j\right\}  
\end{array}\right. \!\!\!\! \right\}, 
\label{eq:rateoverall}
\end{equation}
where $C_{k,s}^{_{[n]}}$ and $C_{k,1,2}^{_{[n]}}$ denote the achievable rate bounds for decoding signal $x_{fs}$, $s\in\left\{ 1,2\right\}$, and signals $\left\{ x_{f1}, x_{f2}\right\}$ at user $k\in\left\{ i,j\right\}$ employing decoding order $n$, respectively, cf. \eqref{eq:rate-II-1-1} and \eqref{eq:rate-II-2-1}\footnote{Due to limited space, we define $C_{k,s}^{_{[n]}}$ and $C_{k,1,2}^{_{[n]}}$ only for $n=1, 2$ in \eqref{eq:rate-II-1-1} and \eqref{eq:rate-II-2-1}. However, they can be similarly defined in a straightforward manner for $n=3,\ldots,8$ in \eqref{eq:rate-III-1}--\eqref{eq:rate-IV-3-3}.}. If only C2 is present in $\mathcal{R}_{n}(\mathcal{P}_{n})$ in Section~\ref{sub3-1}, as is e.g. the case for UE $i$ in \eqref{eq:rate-II-1-1}, C3 can be added without impacting the rate region by defining $C_{k,1,2}^{_{[n]}} = C_{k,1}^{_{[n]}} + C_{k,2}^{_{[n]}}$. 

For an efficient system design, we study the Pareto optimal rate tuples on the boundary of the achievable rate region \cite{Boyd2004Convex}. Assume that $\mathbf{r}^{*}\in\mathcal{R}  (\mathcal{P}) $ is Pareto optimal. Then, a rate tuple satisfying $\mathbf{r}\succeq\mathbf{r}^{*}$ is feasible, i.e., $\mathbf{r}\in\mathcal{R}  (\mathcal{P})$, only if $\mathbf{r}=\mathbf{r}^{*}$. That is, it is impossible to improve one transmission rate without decreasing at least one of the other transmission rates.

On the other hand, to maximize the performance, power and rate have to be jointly optimized. Unfortunately, with adaptive power allocation, the derivation of the Pareto optimal rate tuples is not straightforward. In particular,  as the union of convex sets is not necessarily convex, the rate regions $\mathcal{R}_{n} (\mathcal{P}_{n})$ and $\mathcal{R} (\mathcal{P} )$  are in general nonconvex. To characterize $\mathbf{r}^{*}$ for the nonconvex rate region $\mathcal{R}  (\mathcal{P} )$, we consider the following rate maximization problem\footnote{As $\mathcal{R} (\mathcal{P})$ is nonconvex, weighted sum rate maximization is not applicable for computing $\mathbf{r}^{*}$ \cite[Ch. 4.7]{Boyd2004Convex}.} \cite{Zhang10:RateProfile}:
\begin{alignat}{1}
\textrm{P0:}\;\max_{\mathbf{r}\in\mathcal{R},\,\mathbf{p}\in\mathcal{P},\,r_{\Sigma}\ge0} & r_{\Sigma}\\
\textrm{s.t.}\quad\quad\; & \textrm{C4:}\,r_{k,s}\ge\nu_{k,s}r_{\Sigma},\; k\in\left\{ i,j\right\} ,\,s\in\left\{ 1,2\right\} ,\nonumber 
\end{alignat}
where $\nu_{k,s}\in[0,1]$ are constants fulfilling $\sum_{k\in\left\{ i,j\right\}} \sum_{s\in\left\{ 1,2\right\}}\nu_{k,s}=1$\footnote{For Cases \mbox{II} and \mbox{III}, we have $\nu_{i,1} = 0$ and $\nu_{j,1} = 0$, respectively.}. Problem P0 is not jointly convex with respect to (w.r.t.) $\mathbf{r}$ and $\mathbf{p}$ as $C_{k,s}^{_{[n]}} (\mathbf{p})$ and $C_{k,1,2}^{_{[n]}} (\mathbf{p})$ in C2 and C3 may be nonconvex, cf. \eqref{eq:rateoverall}. {{Hence, the (globally) optimal solution of Problem P0 is not directly available. Nevertheless, we notice that the objective function of Problem P0 is monotonically increasing with respect to $r_{\Sigma}$ and that problem P0 is always feasible for any $r_{\Sigma}\in[0,r_{\Sigma}^{*}]$, where $r_{\Sigma}^{*}$ is the optimal value of problem P0. In the following, we show that, because of these properties, Problem P0 can be solved by evaluating a sequence of convex problems \cite{Zhang10:RateProfile}.}} In particular, for given $r_{\Sigma}$, the following feasibility problems can be defined 
\begin{alignat}{1}
\textrm{P0}(n)\mathrm{\textrm{:}}\;\max_{\mathbf{p}\in\mathcal{P}_{n}}\;\; & 1  \label{eq26:feas} \\
\mathrm{\textrm{s.t.}}\;\;\; & \textrm{C5:}\;C_{k,s}^{_{[n]}}(\mathbf{p})\ge\nu_{k,s}r_{\Sigma},\, k\in\left\{ i,j\right\} ,\,s\in\left\{ 1,2\right\} ,\nonumber \\
 & \textrm{C6:}\;C_{k,1,2,}^{_{[n]}}(\mathbf{p})\ge(\nu_{k,1}+\nu_{k,2})r_{\Sigma},\, k\in\left\{ i,j\right\} ,\nonumber 
\end{alignat}
for decoding order $n \in \mathcal{N}$, where constraints C5 and C6 are equivalent reformulations of C2--C4. The optimal value of P0, $r_{\Sigma}^{*}$, can be found iteratively by employing Algorithm~\ref{alg1}. In particular, in each iteration, the feasibility problems P0$(n)$, $n \in \mathcal{N}$, are solved for a given $r_{\Sigma}$, cf. line~\ref{alg1:line4}. We have $r_{\Sigma}^{*}\ge r_{\Sigma}$ if problem P0$(n)$ is feasible for some $n$, i.e., $r_{\Sigma}$ is a lower bound on $r_{\Sigma}^{*}$, and $r_{\Sigma}^{*}\le r_{\Sigma}$ otherwise, i.e., $r_{\Sigma}$ is an upper bound on $r_{\Sigma}^{*}$. Hence, a bisection search can be applied to iteratively update the value of $r_{\Sigma}$ until the gap between the lower and the upper bounds vanishes, whereby $r_{\Sigma}^*$ is obtained. Moreover, efficient convex optimization algorithms can be employed \cite{Boyd2004Convex} in line~\ref{alg1:line4} of Algorithm~\ref{alg1}. This is because although C5 and C6 are linear fractional constraints of the form $\log_{2}\left(1+\frac{\mathbf{a}^{T}\mathbf{p}}{\mathbf{b}^{T}\mathbf{p}+1}\right)\ge c$ for $\mathbf{a}, \mathbf{b}\in \mathbb{R}_+^4$ and $c \in \mathbb{R}_+$, they can be transformed into equivalent convex constraints of the form $\left(\mathbf{a}-\left(2^{c}-1\right)\mathbf{b}\right)^{T}\mathbf{p}\ge2^{c}-1$ such that an equivalent convex formulation of problem P0$(n)$ is obtained. Therefore, for given $\nu_{k,s}$, $k\in\{i,j\}$, $s\in\{1,2\}$, the obtained solution of P0 defines a Pareto optimal rate tuple with $r_{k,s}^* = \nu_{k,s} r_{\Sigma}^*$. The remaining Pareto optimal points are  obtained by varying $\nu_{k,s}$ \cite{Zhang10:RateProfile}. {{Note that, given an initial search range $[LB, UB]$, the computational complexity of Algorithm~\ref{alg1} is $\mathcal{O}\left(\log_{2}\frac{UB-LB}{\epsilon}\right)$ \cite[Ch. 4.2.5]{Boyd2004Convex}, where $\mathcal{O}\left(\cdot\right)$ is the big-O notation and $\epsilon > 0$ is the tolerance.}}

\begin{algorithm}[t]
\protect\caption{\textcolor{black}{Bisection search for $r_{\Sigma}^*$.} }

\label{alg1} 
\small{
\begin{algorithmic}[1]
\STATE \textbf{initialization}: {Set} $LB \rightarrow 0$, $UB \rightarrow C \big( \frac{P}{\alpha_j} \big) + C \big( \frac{P}{\alpha_i} \big)$, and tolerance $\epsilon$;  

\REPEAT 
\STATE $r_{\Sigma} \rightarrow  (LB+UB)/2   $;
\STATE  Solve feasibility problem \eqref{eq26:feas} for $r_{\Sigma}$ and each $n \in \mathcal{N}$;\label{alg1:line4}
\IF{\eqref{eq26:feas} is feasible for some $n$}{\STATE $LB \rightarrow r_{\Sigma}$;} \ELSE{\STATE $UB \rightarrow r_{\Sigma}$;} \ENDIF
\UNTIL{$UB - LB < \epsilon$.} \label{alg1:line17} 
\end{algorithmic} 
}
\end{algorithm}

\section{\label{sec4}Rate and Power Allocation for Delivery Time Minimization}
 
In this section, cache-aided NOMA is exploited for fast video streaming. To this end, we first formulate a joint decoding order, transmit power, and rate optimization problem for minimization of the delivery time. Then, the optimal solutions are characterized as functions of the cache status, the requested file sizes, and the channel conditions across UEs. 

\subsection{\label{sub4-1}Optimization Problem Formulation}

Let $T$ be the time required to complete the delivery of the requested files. We have 
\begin{equation}
T=\max_{k\in\left\{ i,j\right\} ,\,s\in\left\{ 1,2\right\}}\;\frac{\beta_{k,s}}{r_{k,s}},\label{eq:del-time}
\end{equation}
for $\mathbf{r}\in\mathcal{R}$, where $\beta_{k,1}\triangleq(\overline{c}_{f}-c_{kf})V_{f}$ and $\beta_{k,2}\triangleq(1-\overline{c}_{f})V_{f}$, $(k,f)\in\left\{ (i,A),(j,B)\right\}$, denote the effective volume of data to be delivered to UE $k$. To avoid trivial results, we assume throughout this section that $\beta_{k,1}+\beta_{k,2}>0$, $\forall k\in\left\{ i,j\right\} $, i.e., both UEs request some video data that is not cached\footnote{If $\beta_{k,1}+\beta_{k,2} = 0$, we have $p_{k,1}=p_{k,2}=0$ and $r_{k,1}=r_{k,2}=0$.}. Consequently, the delivery time optimization problem is formulated as 
\begin{alignat}{1}
\textrm{P1:}\;\min_{\mathbf{r}\in\mathcal{R},\;\mathbf{p}\in\mathcal{P},\;T\ge0}\quad & T\\
\textrm{s.t.}\qquad\quad\; & \textrm{C7:}\;r_{ks}T\ge\beta_{k,s},\; k\in\left\{ i,j\right\} ,\,s\in\left\{ 1,2\right\}, \nonumber 
\end{alignat}
where constraint C7 ensures that file delivery completes no later than time $T$. 

Problem P1 is in general nonconvex as the capacity bound functions in C2 and C3 in \eqref{eq:rateoverall} are not jointly convex w.r.t. $\mathbf{r}$ and $\mathbf{p}$, and C7 is bilinear. This type of problem is usually NP hard. However, based on necessary conditions that the optimal solution of Problem P1 has to fulfill, the optimal solution can be analytically derived. 

\subsection{\label{sub4-2}Optimal Solution of Problem P1}
For the optimal decoding order, the optimal power and rate allocation have to satisfy the conditions specified in Lemma~\ref{lem1}.
\begin{lem}
\emph{\label{lem1} Assume that decoding order $n$ is optimal. Then, the optimal power and rate allocation policy, denoted by $p_{k,s}^*$ and $r_{k,s}^*$, $k\in \{i,j\}$, $s\in \{1,2\}$, that solves Problem P1 necessarily fulfills:
\begin{alignat}{1}
&\beta_{i,s'}r_{j,s}^* =\beta_{j,s}r_{i,s'}^*,\quad s,s'\in\left\{ 1,2\right\}, \label{eq:opt_cond1}\\
&r_{k,s}^{*}\left(\mathbf{p}^*\right) \!=\! 
\begin{cases}
\frac{\beta_{k,s}}{\beta_{k,1}+\beta_{k,2}}C_{k,1,2}^{_{[n]}}, & {\text{if $C_{k,1}^{_{[n]}} \!+\! C_{k,2}^{_{[n]}} \!>\! C_{k,1,2}^{_{[n]}}$,}}\\
C_{k,s}^{_{[n]}}, & {\text{otherwise}},  
\end{cases}
 \label{eq:opt_cond2} \\
& p_{i,1}^*+p_{i,2}^*+p_{j,1}^*+p_{j,2}^* =P. \label{eq:opt_cond3}
\end{alignat} }
\end{lem}
\begin{IEEEproof} 
The proof requires checking whether the inequality constraints, e.g., C7, hold with equality at optimum. To this end, a proof by contradiction similar to that for Proposition~3 in \cite{Xiang17TWC:Untrusted} can be constructed. Due to the limited space, the details are omitted here. 
\end{IEEEproof}

According to Lemma~\ref{lem1}, the optimal rate allocations $r_{k,s}^*$ are proportional to the effective delivery sizes $\beta_{k,s}$ in \eqref{eq:opt_cond1}, i.e., C7 is active. Especially, if $\beta_{k,s}=0$, \eqref{eq:opt_cond1} implies $r_{k,s}=0$ for $k\in\left\{ i,j\right\} $, $s\in\left\{ 1,2\right\}$.  Moreover, for any feasible power allocation, the rate region $\mathcal{R}_{n}$, cf. \eqref{eq:rateoverall}, reduces to a polyhedron. Consequently, \eqref{eq:opt_cond2} ensures that the optimal rate tuple is located on the dominant face\footnote{For a polyhedron, any point that lies outside the dominant face is dominated elementwise by a point on the dominant face \cite{Tse2005Fundamentals}.} of $\mathcal{R}_{n}$ \cite[pp. 231]{Tse2005Fundamentals}. Finally, \eqref{eq:opt_cond3} indicates that the optimal power allocation utilizes the maximum possible transmit power.

\begin{table*}
\renewcommand{\arraystretch}{1.6}
\centering
\caption{\label{tab1}Optimal interference-plus-noise power, $I_{k,s}^*$, and decoding order for Case \mbox{I}. Optimal rate allocation is given by $r_{k,s}^* = C\left( {p_{k,s}^*} / {I_{k,s}^*}\right)$, where $p_{k,s}^*$ is provided in Table~\ref{tab2}.}
\vspace{-.2cm}
\footnotesize
\begin{tabular}{c c l}
\hline
\hline  
 Index  & {Optimal value of $I_{i,1}^*, I_{i,2}^*, I_{j,1}^*, I_{j,2}^*$ } & {\qquad Optimal decoding order} \tabularnewline
\hline 
{$n=1$ } & {${p_{i,2}^*+p_{j,2}^*+\alpha_{i}}, \, {\alpha_{i}}, \,
\begin{Bmatrix}
{p_{j,2}^*+p_{i,2}^*+\alpha_{j}}, \, {p_{i,2}^*+\alpha_{j}} \\
{ p_{i,2}^*+\alpha_{j}}, \, {p_{i,2}^*+ p_{j,1}^* + \alpha_{j}} 
\end{Bmatrix}  $}
 & {$i\overset{_{(1)}}{\to}x_{A1}\overset{_{(2)}}{\to}x_{B2}\overset{_{(3)}}{\to}x_{A2}$, $
j\overset{_{(1)}}{\to} \!\!
\begin{cases}
x_{B1}\overset{_{(2)}}{\to}x_{B2} \\
x_{B2}\overset{_{(2)}}{\to}x_{B1} 
\end{cases} $} \tabularnewline
{$n=2$} & {$ {p_{i,2}^*+p_{j,2}^*+\alpha_{i}}, \, {p_{j,2}^*+\alpha_{i}}, \, {\alpha_{j}}, \, {p_{i,2}^*+p_{j,1}^*+\alpha_{j}}$ }  & {$i\overset{_{(1)}}{\to}x_{A1}\overset{_{(2)}}{\to}x_{A2}$,} {$j\overset{_{(1)}}{\to}x_{A2}\overset{_{(2)}}{\to}x_{B2}\overset{_{(3)}}{\to}x_{B1}$}  \tabularnewline
{$n=3$} & {$
\begin{Bmatrix}
{p_{i,2}^*+\alpha_{i}}, \, {\alpha_{i}}, \\
{\alpha_{i}},\, {p_{i,1}^*+\alpha_{i}}, 
\end{Bmatrix}
\,{p_{i,2}^*+p_{j,2}^*+\alpha_{j}}, \, {p_{i,2}^*+\alpha_{j}}$ } & {$i\overset{_{(1)}}{\to}x_{B2}\overset{_{(2)}}{\to}
\begin{cases}
x_{A1}\overset{_{(3)}}{\to} x_{A2}, \\
x_{A2}\overset{_{(3)}}{\to} x_{A1}, 
\end{cases}
\!\! j\overset{_{(1)}}{\to}x_{B1}\overset{_{(2)}}{\to}x_{B2}$}\tabularnewline
{$n=4$} & {$ {p_{j,2}^*+\alpha_{i}}, \, {p_{i,1}^*+p_{j,2}^*+\alpha_{i}},\, {p_{i,2}^*+p_{j,2}^*+\alpha_{j}},\, {\alpha_{j}}$} & {$i\overset{_{(1)}}{\to}x_{A2}\overset{(2)}{\to}x_{A1}$}, {$j\overset{_{(1)}}{\to}x_{B1}\overset{_{(2)}}{\to}x_{A2}\overset{_{(3)}}{\to}x_{B2}$} \tabularnewline
{$n=5$} & {$ {p_{i,2}^*+p_{j,2}^*+\alpha_{i}},\, {\alpha_{i}},\, {p_{i,2}^*+p_{j,2}^*+\alpha_{j}},\, {p_{i,2}^*+\alpha_{j}}$} & {$i\overset{_{(1)}}{\to}x_{A1}\overset{_{(2)}}{\to}x_{B2}\overset{_{(3)}}{\to}x_{A2}$}, {$j\overset{_{(1)}}{\to}x_{B1}\overset{_{(2)}}{\to}x_{B2}$} \tabularnewline
{$n=6$} & {$
\begin{Bmatrix}
 {p_{i,2}^*+\alpha_{i}},\, {\alpha_{i}}, \\
 {\alpha_{i}},\, {p_{i,1}^*+\alpha_{i}},
\end{Bmatrix}
 \, {p_{i,2}^*+\alpha_{j}},\, {p_{j,1}^*+p_{i,2}+\alpha_{j}}$ } & {$i\overset{_{(1)}}{\to}x_{B2}\overset{_{(2)}}{\to}
\begin{cases}
x_{A1}\overset{_{(3)}}{\to} x_{A2}, \\
x_{A2}\overset{_{(3)}}{\to} x_{A1}, 
\end{cases}
\!\! j\overset{_{(1)}}{\to}x_{B2}\overset{_{(2)}}{\to}x_{B1}$} \tabularnewline
{$n=7$} & {$ {\alpha_{i}+p_{j,2}^* \Delta },\, {p_{i,1}^*+p_{j,2}^*+\alpha_{i}},\, {\alpha_{j}},\,  {p_{i,2}^*+p_{j,1}^*+\alpha_{j}}$}  & {$i\overset{_{(1)}}{\to}x_{A2}\overset{_{(2)}}{\to} \!\! \begin{cases}
x_{A1}, \textrm{ if } \Delta = 1,\\
x_{B2}\overset{_{(3)}}{\to}x_{A1}, \textrm{ otherwise,}
\end{cases}
j\overset{_{(1)}}{\to}x_{B2}\overset{_{(2)}}{\to}x_{B1}$} \tabularnewline
{$n=8$} & {$ {p_{j,2}^*+\alpha_{i}},\, {p_{i,1}^*+p_{j,2}^*+\alpha_{i}},\,
\begin{Bmatrix}
 {p_{j,2}^*+\alpha_{j}},\, {\alpha_{j}} \\
 {\alpha_{j}},\, {p_{j,1}^*+\alpha_{j}}
\end{Bmatrix}
$}  &  {$i\overset{_{(1)}}{\to}x_{A2}\overset{_{(2)}}{\to}x_{A1}$}, {$j\to x_{A2}\overset{_{(2)}}{\to} \begin{cases}
x_{B1}\overset{_{(3)}}{\to}x_{B2} \\
x_{B2}\overset{_{(3)}}{\to}x_{B1} 
\end{cases}
$}  \tabularnewline 
\hline
\hline 
\end{tabular}
\end{table*}

Lemma~\ref{lem1} provides the set of equations required for solving the optimal power and rate allocation problem for a given decoding order. 
By solving \eqref{eq:opt_cond1}--\eqref{eq:opt_cond3} in Lemma~\ref{lem1}, the optimal solution of Problem P1 for Case \mbox{I} can be obtained in closed form, cf. Proposition~\ref{prop5}. 

\begin{prop}
{\emph{\label{prop5} For Case \mbox{I}, the optimal rate and power allocation for decoding order $n\in \{1,\ldots,8\}$ are given by 
\begin{equation}
r_{k,s}^* =  C \Big( \frac{p_{k,s}^*}{I_{k,s}^*} \Big), \quad \textrm{and} \quad  p_{k,s}^* = \left( \gamma_n^{\beta_{k,s}} -1 \right)  I_{k,s}^*, 
\label{eq:optpwr}
\end{equation}
respectively, where $I_{k,s}^*$, $k\in \{i,j\}$, $s\in \{1,2\}$, denotes the residual interference-plus-noise power for decoding signal $x_{k,s}$ if $r_{k,s}^*$ and $ p_{k,s}^*$ are employed, and $\gamma_n$ is chosen such that \eqref{eq:opt_cond3} is fulfilled. Moreover, the optimal delivery time is given by 
\begin{equation}
T_{\lyxmathsym{\mbox{I}}}^{*} =\frac{1}{\log_{2}\gamma_{\lyxmathsym{\mbox{I}}}^{*}},\textrm{ with }\gamma_{\lyxmathsym{\mbox{I}}}^{*}=\max_{n\in \{1,\ldots,8\}} \gamma_n. 
\label{eq:deltime}
\end{equation}
The decoding order and the respective powers and rates that achieve $T_{\lyxmathsym{\mbox{I}}}^{*}$ are optimal. The expressions for the optimal interference-plus-noise and transmit powers for all feasible decoding orders are provided in Tables~\ref{tab1} and \ref{tab2}, respectively. }}
\end{prop}
\begin{IEEEproof}
We show the proof only for $n=1$, i.e., for $\mathbf{r}\in\mathcal{R}_{\text{\mbox{I}},1}$ (cf. Proposition \ref{prop1}), but the same approach can be used to prove the results for the other values of $n$. According to \eqref{eq:opt_cond2},  if $x_{B2}$ is decoded after $x_{B1}$, the optimal rate allocation is as follows,
\vspace{-.2cm}
\begin{alignat}{1}
r_{i,1}^{*} &=C_{i,1}^{_{[1]}},\, r_{j,1}^{*} \!=\! \tfrac{\beta_{j,1}}{\beta_{j,1}+\beta_{j,2}}C_{j,1,2}^{_{[1]}} =C\left(\tfrac{p_{j,1}^{*}}{p_{j,2}^{*}+p_{i,2}^{*}+\alpha_{j}}\right),\\
r_{i,2}^{*} &=C_{i,2}^{_{[1]}},\, r_{j,2}^{*} \!=\! \tfrac{\beta_{j,2}}{\beta_{j,1}+\beta_{j,2}}C_{j,1,2}^{_{[1]}} =C\left(\tfrac{p_{j,2}^{*}}{p_{i,2}^{*}+\alpha_{j}}\right).\nonumber 
\end{alignat}
Otherwise, the optimal rate allocation is given by 
\vspace{-.2cm}
\begin{alignat}{1}
r_{i,1}^{*} &=C_{i,1}^{_{[1]}},\, r_{j,1}^{*} =\tfrac{\beta_{j,1}}{\beta_{j,1}+\beta_{j,2}}C_{j,1,2}^{_{[1]}} =C\left(\tfrac{p_{j,1}^{*}}{p_{i,2}^{*}+\alpha_{j}}\right),\\
r_{i,2}^{*} &=C_{i,2}^{_{[1]}},\, r_{j,2}^{*} =\tfrac{\beta_{j,2}}{\beta_{j,1}+\beta_{j,2}}C_{j,1,2}^{_{[1]}} =C\left(\tfrac{p_{j,2}^{*}}{p_{i,2}^{*}+p_{j,1}^{*}+\alpha_{j}}\right).\nonumber 
\end{alignat}
In both cases, we have $r_{k,s}^* = C \left( \tfrac{p_{k,s}^*}{I_{k,s}^*} \right)$. Meanwhile, according to \eqref{eq:opt_cond1}, the optimal power allocation fulfills 
\begin{equation}
\tfrac{\beta_{i,1}}{r_{i,1}^{*}}=\tfrac{\beta_{i,2}}{r_{i,2}^{*}}=\tfrac{\beta_{j,1}}{r_{j,1}^{*}} =\tfrac{\beta_{j,2}}{r_{j,2}^{*}} = \tfrac{1}{\log_2 \gamma_1},
\end{equation}
where $\gamma_{1}$ is chosen such that \eqref{eq:opt_cond3} holds. Hence, $p_{k,s}^* = \big( \gamma_n^{\beta_{k,s}} -1 \big)  I_{k,s}^*$. By solving the above system of equations, the optimal power allocation for $n=1$ can be obtained as in Table~\ref{tab2}. This completes the proof.
\end{IEEEproof}

\begin{table*}
\renewcommand{\arraystretch}{1.7}
\centering\protect\protect\caption{\label{tab2}Optimal power allocation, $p_{k,s}^*$, for Case \mbox{I}.}
\vspace{-.3cm}
\footnotesize
\begin{tabular}{c c}
\hline
\hline  
 Index  & Optimal power allocation $p_{i,1}^*, p_{i,2}^*, p_{j,1}^*, p_{j,2}^*$  \tabularnewline
\hline
{$n=1$ } & {$(\gamma_{1}^{\beta_{i,1}}-1)(\alpha_{i}\gamma_{1}^{\beta_{i,2}}+p_{j,2}^*), \; \alpha_{i}(\gamma_{1}^{\beta_{i,2}}-1),$} 
{$ \begin{Bmatrix} 
\gamma_{1}^{\beta_{j,2}}(\gamma_{1}^{\beta_{j,1}} \!-\! 1)(\alpha_{i}\gamma_{1}^{\beta_{i,2}} \!+\! \alpha_{j} \!-\!\alpha_{i}), \; (\gamma_{1}^{\beta_{j,2}} \!-\! 1) (\alpha_{i}\gamma_{1}^{\beta_{i,2}} \!+\! \alpha_{j} \!-\! \alpha_{i}) \\
(\gamma_{1}^{\beta_{j,1}} \!-\! 1) (\alpha_{i}\gamma_{1}^{\beta_{i,2}} \!+\! \alpha_{j} \!-\! \alpha_{i}), \; \gamma_{1}^{\beta_{j,1}}(\gamma_{1}^{\beta_{j,2}} \!-\! 1)(\alpha_{i}\gamma_{1}^{\beta_{i,2}} \!+\! \alpha_{j} \!-\!\alpha_{i})
\end{Bmatrix}
$} \tabularnewline
{$n=2$} & {$\frac{\gamma_{2}^{\beta_{i,2}} (\gamma_{2}^{\beta_{i,1}}-1 )}{\gamma_{2}^{\beta_{i,2}}-1}p_{i,2}^*, \quad  \frac{ (\gamma_{2}^{\beta_{i,2}}-1 ) (\alpha_{i}+\alpha_{j}\gamma_{2}^{\beta_{j,1}+\beta_{j,2}}-\alpha_{j}\gamma_{2}^{\beta_{j,1}} )}{\gamma_{2}^{\beta_{j,2}}+\gamma_{2}^{\beta_{i,2}}-\gamma_{2}^{\beta_{i,2}+\beta_{j,2}}},$ } 
$\alpha_{j} (\gamma_{2}^{\beta_{j,1}}-1 ), \quad \frac{(\gamma_{2}^{\beta_{j,2}}-1) (\alpha_{i}\gamma_{2}^{\beta_{i,2}}+\alpha_{j}\gamma_{2}^{\beta_{j,1}}-\alpha_{i})}{\gamma_{2}^{\beta_{j,2}}+\gamma_{2}^{\beta_{i,2}}-\gamma_{2}^{\beta_{i,2}+\beta_{j,2}}}$  \tabularnewline
{$n=3$} & {$\begin{Bmatrix}
\alpha_{i}\gamma_{3}^{\beta_{i,2}} (\gamma_{3}^{\beta_{i,1}}-1 ), \quad \alpha_{i} (\gamma_{3}^{\beta_{i,2}}-1 ), \\
\alpha_{i} (\gamma_{3}^{\beta_{i,1}}-1 ),\quad \alpha_{i}\gamma_{3}^{\beta_{i,1}} (\gamma_{3}^{\beta_{i,2}}-1 ), 
\end{Bmatrix}
$ }  
{$(\gamma_{3}^{\beta_{j,1}}-1 ) (p_{i,2}^*+p_{j,2}^*+\alpha_{j} ), \quad \alpha_{i}\gamma_{3}^{\beta_{i,1}+\beta_{i,2}} (\gamma_{3}^{\beta_{j,2}}-1 )$ } \tabularnewline
{$n=4$} & {$(\gamma_{4}^{\beta_{i,1}} \!-\! 1) (\alpha_{j}\gamma_{4}^{\beta_{j,2}} \!-\! \alpha_{j} \!+\! \alpha_{i} ), \; \gamma_{4}^{\beta_{i,1}} (\gamma_{4}^{\beta_{i,2}} \!-\! 1 ) (\alpha_{j}\gamma_{4}^{\beta_{j,2}} \!-\!\alpha_{j} \!+\! \alpha_{i} ),$}  
{$ (\gamma_{4}^{\beta_{j,1}}-1 ) (p_{i,2}^*+p_{j,2}^*+\alpha_{j} ), \quad \alpha_{j} (\gamma_{4}^{\beta_{j,2}}-1 )$}   \tabularnewline
{$n=5$} & {$ (\gamma_{5}^{\beta_{i,1}}-1)(\alpha_{i}\gamma_{5}^{\beta_{i,2}}+p_{j,2}^*), \quad \alpha_{i}(\gamma_{5}^{\beta_{i,2}}-1),$} 
{$ \gamma_{5}^{\beta_{j,2}}(\gamma_{5}^{\beta_{j,1}} \!-\! 1)(\alpha_{i}\gamma_{5}^{\beta_{i,2}} \!+\! \alpha_{j} \!-\!\alpha_{i}), \; (\gamma_{5}^{\beta_{j,2}} \!-\! 1) (\alpha_{i}\gamma_{5}^{\beta_{i,2}} \!+\! \alpha_{j} \!-\! \alpha_{i}) $}  \tabularnewline
{$n=6$} & {$\begin{Bmatrix} 
\alpha_{i}\gamma_{6}^{\beta_{i,2}} (\gamma_{6}^{\beta_{i,1}}-1), \quad \alpha_{i} (\gamma_{6}^{\beta_{i,2}}-1),\\
\alpha_{i} (\gamma_{6}^{\beta_{i,1}}-1), \quad \alpha_{i}\gamma_{6}^{\beta_{i,1}} (\gamma_{6}^{\beta_{i,2}}-1),
\end{Bmatrix}
$ }   
{$(\gamma_{6}^{\beta_{j,1}}-1) (p_{i,2}^*+\alpha_{j}), \quad (\gamma_{6}^{\beta_{j,2}}-1) (p_{j,1}^*+p_{i,2}^*+\alpha_{j})$} \tabularnewline
{$n=7$} & {$(\gamma_{7}^{\beta_{i,1}}-1) (\alpha_{i}+p_{j,2}^* \Delta ), \; (\gamma_{7}^{\beta_{i,2}}-1) (p_{i,1}^*+p_{j,2}^*+\alpha_{i}),$}  
{$\alpha_{j} (\gamma_{7}^{\beta_{j,1}}-1), \; (\gamma_{7}^{\beta_{j,2}}-1) (p_{i,2}^*+p_{j,1}^*+\alpha_{j})$}   \tabularnewline
{$n=8$} & {$(\gamma_{8}^{\beta_{i,1}}-1) (p_{j,2}^*+\alpha_{i}), \quad (\gamma_{8}^{\beta_{i,2}}-1) (p_{i,1}^*+p_{j,2}^*+\alpha_{i}),$}  
{$\begin{Bmatrix}
\alpha_{j}\gamma_{8}^{\beta_{j,2}} (\gamma_{8}^{\beta_{j,1}}-1), \quad \alpha_{j} (\gamma_{8}^{\beta_{j,2}}-1) \\
\alpha_{j} (\gamma_{8}^{\beta_{j,1}}-1), \quad \alpha_{j}\gamma_{8}^{\beta_{j,1}} (\gamma_{8}^{\beta_{j,2}}-1)
\end{Bmatrix} $} \tabularnewline
\hline
\hline
\end{tabular}
\end{table*}

\begin{rem}
From Table~\ref{tab2}, we observe that for decoding order $n$, $p_{k,s}^*$ increases exponentially with the effective volume of delivery data $\beta_{k,s}$ and linearly with the effective channel gain $\alpha_k$ of UE $k$. However, as the interference-plus-noise power $I_{k,s}^*$ of UE $k$ may depend on the cache and channel statuses of the other UE, the specific value of $p_{k,s}^*$ has to be calculated according to the cache and channel statuses of both UEs. On the other hand, from Table~\ref{tab1}, we can observe that $I_{k,s}^* \ge I_{k',s'}^*$ whenever $x_{k,s}$ is decoded before $x_{k',s'}$ for any $k, k'\in \{i,j\}$, $s, s'\in \{1,2\}$. For example, for decoding order $n=1$, we have $I_{i,1}^*  \ge I_{j,2}^* \ge  I_{i,2}^* $ if $i\overset{(1)}{\to}x_{A1}\overset{(2)}{\to}x_{B2}\overset{(3)}{\to}x_{A2}$ and $I_{j,2}^* \ge I_{j,1}^* $ ($I_{j,2}^* \le I_{j,1}^*$)  if $j \overset{(1)}{\to} x_{B2}\overset{(2)}{\to}x_{B1}$ ($j \overset{(1)}{\to} x_{B1}\overset{(2)}{\to}x_{B2}$), respectively. Hence, by selecting the decoding order with $x_{Bs}$, $s\in\{1,2\}$, decoded last, the weak user can achieve a high delivery rate and a short delivery time. Furthermore, owing to the SIC decoding condition, we need to search only $8$, rather than $24$, feasible decoding orders in the worst case for determining the optimal decoding order, which substantially reduces the computational complexity.
\end{rem}

\begin{rem}
We note that the computational complexity of cache-aided NOMA is dominated by SIC as CIC does not require decoding. According to Table~\ref{tab1}, at the strong UE, cache-aided NOMA and conventional NOMA have similar computational complexity except that the former requires the interfering and the desired signals to be decoded in multiple SIC stages. On the other hand, at the weak UE, the computational complexity of cache-aided NOMA is  higher than that of conventional NOMA when cache-aided NOMA enables SIC.
\end{rem}

Meanwhile, the optimal delivery times for the cache configurations of Cases \mbox{II}--\mbox{IV} are given in Corollary~\ref{cor2}. 

\begin{cor}
{\emph{\label{cor2} For Cases \mbox{II} and \mbox{III}, the optimal delivery time is given by 
\vspace{-.2cm}
\begin{alignat}{1}
T_{\lyxmathsym{\mbox{II}}}^{*} &=\frac{1}{\log_{2}\gamma_{\lyxmathsym{\mbox{II}}}^{*}},\textrm{ with }\gamma_{\lyxmathsym{\mbox{II}}}^{*}=\max_{n\in \{1,2\}} \gamma_n, \\
T_{\lyxmathsym{\mbox{III}}}^{*} &=\frac{1}{\log_{2}\gamma_{\lyxmathsym{\mbox{III}}}^{*}},\textrm{ with }\gamma_{\lyxmathsym{\mbox{III}}}^{*}=\max_{n\in \{3,4,5\}} \gamma_n.
\end{alignat}
The decoding order and the respective power and rate allocations that minimize the delivery time are obtained as those for $n\in \{1,\ldots,5\}$ in Tables~\ref{tab1} and~\ref{tab2} except that $r_{i,1}^*=0$, $p_{i,1}^*=0$ for $n\in \{1,2\}$ and $r_{j,1}^*=0$, $p_{j,1}^*=0$ for $n\in\{3,4,5\}$. For Case \mbox{IV},  the optimal delivery time is given by $T_{\lyxmathsym{\mbox{IV}}}^{*}=\max\left\{ \frac{\beta_{i,2}}{r_{i,2}^{*}},\frac{\beta_{j,2}}{r_{j,2}^{*}}\right\}$, with optimal rate allocation $r_{i,2}^{*}=C\left(\frac{p_{i,2}^{*}}{\alpha_{i}}\right)$ and $r_{j,2}^{*}=C\left(\frac{p_{j,2}^{*}}{p_{i,2}^{*}+\alpha_{j}}\right)$, and the optimal power allocation $(p_{i,2}^{*},p_{j,2}^{*})$ is obtained from
\vspace{-.2cm}
\begin{alignat}{1}
&p_{j,2}^{*}=\left[\left(1+{p_{i,2}^{*}}/{\alpha_{i}}\right)^{\beta_{j,2}/\beta_{i,2}}-1\right]\left(p_{i,2}^{*}+\alpha_{j}\right)\quad\textrm{and} \nonumber \\
& p_{i,2}^{*}+p_{j,2}^{*}=P.
\label{eq:cor2}
\end{alignat}
 }}
\end{cor}
\begin{IEEEproof}
The proof is similar to that for Proposition~\ref{prop5}. 
\end{IEEEproof}
\begin{rem}
{{In Proposition~\ref{prop5} and Corollary~\ref{cor2}, finding the optimal $\gamma_n$ that fulfills \eqref{eq:opt_cond3} involves a one-dimensional search, for which, e.g., the bisection method can be used, similar to Algorithm~\ref{alg1}. Assume that we start searching for $\gamma_{n}$ in the range $[0,P]$. By bisection search, $\mathcal{O}\left(\log_{2}\frac{P}{\epsilon}\right)$ iterations are needed to determine the optimal power allocation $\gamma_{n}$ with tolerance $\epsilon>0$ \cite[Ch. 4.2.5]{Boyd2004Convex}.}} 
\end{rem}

\section{\label{sec5}Performance Evaluation}

In this section, the performance of the proposed cache-aided NOMA scheme is evaluated by simulation. Consider a cell of radius $R=3$~km, where the BS is deployed at the center of the cell and the strong and the weak users, UE $i$ and UE $j$, are uniformly distributed on discs of radii $R_i$ and $R_j$, respectively. For the wireless channel, the 3GPP path loss model for the ``Urban Macro NLOS'' scenario in \cite{3GPP:TR36814} is adopted, and the small-scale fading coefficients are independently and identically distributed (i.i.d.) Rayleigh random variables. The video files have size $V_{A} = V_{B} =$500 MBytes. Moreover, the system has a bandwidth of $5$~MHz and the noise power spectral density is $-$172.6 dBm/Hz. For ease of reference, we define the caching vector $\mathbf{c} \triangleq \left( c_{iA},c_{iB},c_{jA},c_{jB}\right)$. Unless otherwise specified, we set the maximal transmit power at the BS to $P = 36$ dBm, the radii to $R_i = 1$~km and $R_j = 2$~km, and the caching vector to $\mathbf{c} = \left( 0.2, 0.8, 0.8, 0.2\right)$.  

\subsection{\label{sub3-4}Baseline Schemes}

\subsubsection{Baseline 1 (Cache-aided OMA)}

As baseline, we consider time-division multiple access (TDMA) for transmitting the uncached portions of the requested files. In particular, $\tau$ and $1- \tau$ fractions of time are allocated for transmission to UE $i$ and UE $j$, respectively, where $\tau \in [0,1]$. Consequently, the capacity region for all feasible time allocations is given by 
$\mathcal{R}_{\mathrm{OMA}}=  \bigcup_{\tau \in [0,1] }  \big\{ (r_{i},r_{j})\in\mathbb{R}_{+}^{2} \mid 
r_{i}\le\tau C(\frac{P}{\alpha_{i}}),\;r_{j}\le (1-\tau) C(\frac{P}{\alpha_{j}}) \big\}$.
Note that, with Baseline 1, caching facilitates only conventional offloading of the hit cached data.

For Baseline 1, the delivery time for rate allocations $r_{i} =  \tau C(\frac{P}{\alpha_{i}})$ and $r_{j} = (1-\tau) C(\frac{P}{\alpha_{j}})$ is given by  
$T_{\mathrm{OMA}} (\tau)  = \frac{\mu_{i}}{\tau}+\frac{\mu_{j}}{(1-\tau)}$,
where $\mu_{k}=\frac{\beta_{k}}{C(P/\alpha_{k})}$ and $\beta_{k}\triangleq(1-c_{kf})V_{f}$, $(k,f)\in\left\{ (i,A),(j,B)\right\}$, is the effective volume of the delivery data. Moreover, the delivery time minimization problem for Baseline 1 is formulated as 
\vspace{-.2cm}
\begin{alignat}{1}
\textrm{P2:}\quad\min_{\tau \in [0,1]}\quad & T_{\mathrm{OMA}}(\tau). 
\end{alignat}
By solving Problem P2, the optimal time allocation for Baseline 1 is obtained as $\tau^{*}=\frac{\sqrt{\mu_{i}}}{\sqrt{\mu_{i}}+\sqrt{\mu_{j}}}$, and the optimal delivery time is $T_{\mathrm{OMA}}^{*}(\tau^{*})=\left(\sqrt{\mu_{i}}+\sqrt{\mu_{j}}\right)^{2}$.

\subsubsection{Baseline 2 (Conventional NOMA with and without caching)}

If caching is possible, Baseline 2 is a straightforward combination of caching and NOMA, whereby the requested data hit by the cache is offloaded and only the remaining data is transmitted by applying NOMA \cite{Ding17Noma:Caching}. If caching is not possible, Baseline 2 reduces to the conventional NOMA scheme. In both cases, the BS transmits signal $x=\sqrt{p_{i}}x_{A}+\sqrt{p_{j}}x_{B}$ for delivering files $W_A$ and $W_B$, where the feasible set for power allocations $p_{i}$ and $p_{j}$ is $\mathcal{P}_{\mathrm{NOMA}}\triangleq\left\{ (p_{i},p_{j})\in\mathbb{R}_{+}^{2}\mid p_{i}+p_{j}\le P\right\}$. Signals $y_{i}  =h_{i}\left(\sqrt{p_{i}}x_{A}+\sqrt{p_{j}}x_{B}\right)+z_{i}$ and $y_{j} =h_{j}\left(\sqrt{p_{i}}x_{A}+\sqrt{p_{j}}x_{B}\right)+z_{j}$ are received at UEs $i$ and $j$, respectively. Employing SIC, the capacity region $\mathcal{R}_{\mathrm{NOMA}}(\mathcal{P}_{\mathrm{NOMA}})=\bigcup_{\mathbf{p}\in\mathcal{P}_{\mathrm{NOMA}}} \big\{ (r_{i},r_{j}) \mid r_{i}\le C\big(\frac{p_{i}}{\alpha_{i}}\big),\;r_{j}\le C\big(\frac{p_{j}}{p_{i}+\alpha_{j}}\big) \big\}$ is achieved with and without caching \cite{Cover12IT-Elements,Gamal11NIT}, where, at UE $i$, $x_{B}$ is decoded and canceled before decoding $x_{A}$. Moreover, according to Corollary~\ref{cor2}, the optimal delivery time for Baseline 2 is given by $T_{\text{NOMA}}^{*}=\max\left\{ \frac{\beta_{i}}{r_{i}^{*}},\frac{\beta_{j}}{r_{j}^{*}}\right\} $, where the optimal power allocation is obtained as $ p_{i}^{*} = p_{i,2}^{*}$, $p_{j}^{*}= p_{j,2}^{*}$, and $p_{i,2}^{*}$ and $p_{j,2}^{*}$ are given in \eqref{eq:cor2}. For Baseline 2 without cache, we set $c_{iA}=c_{jB}=0$ in the definition of $\beta_{i}$ and $\beta_{j}$. 

\subsection{Simulation Results}

\begin{figure}[t]
\centering
\subfloat[]{\label{fig3a} \includegraphics[width=3.0in]{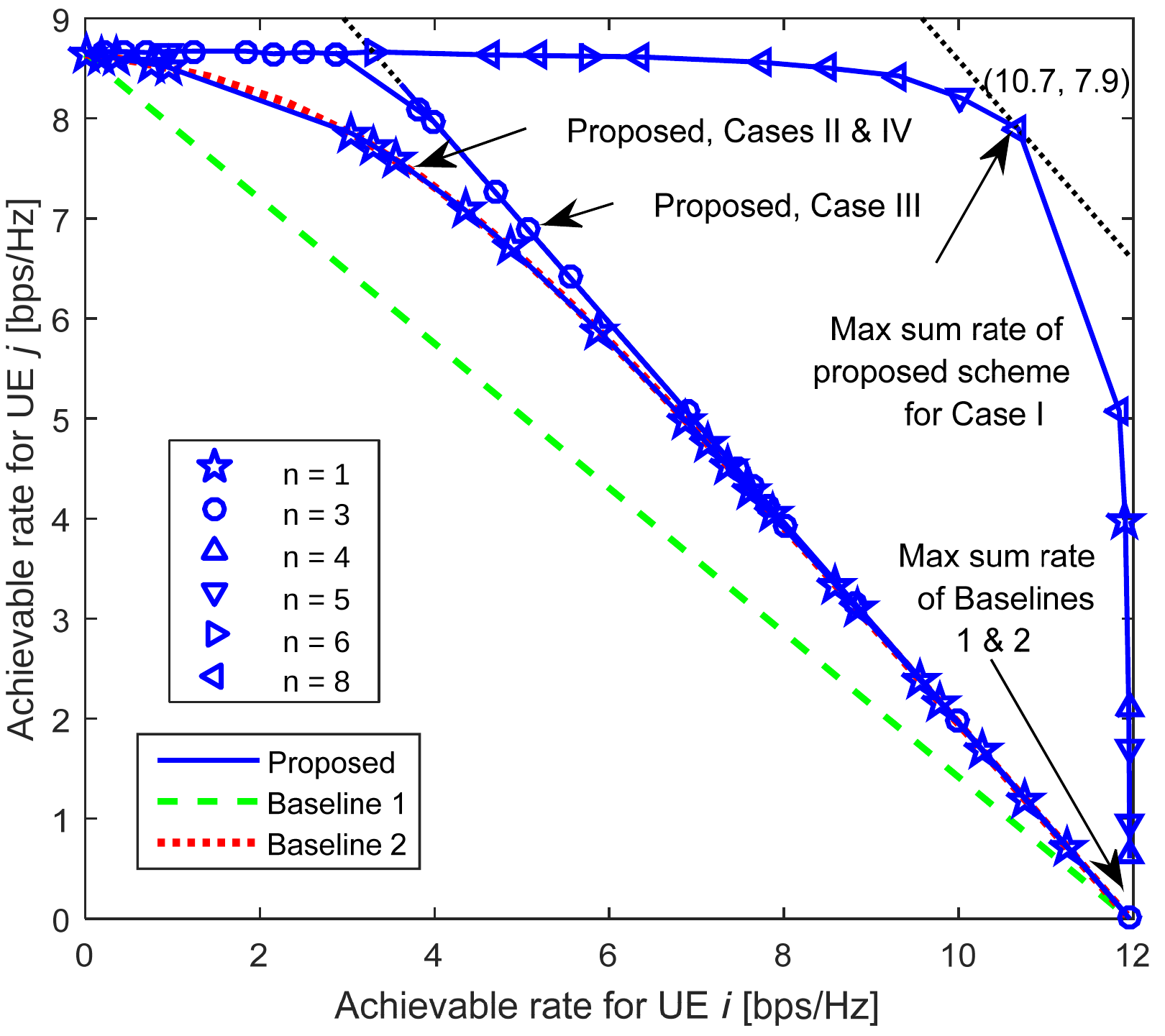}} 

\vspace{-.3cm}
\subfloat[]{\label{fig3b} \includegraphics[width=3.0in]{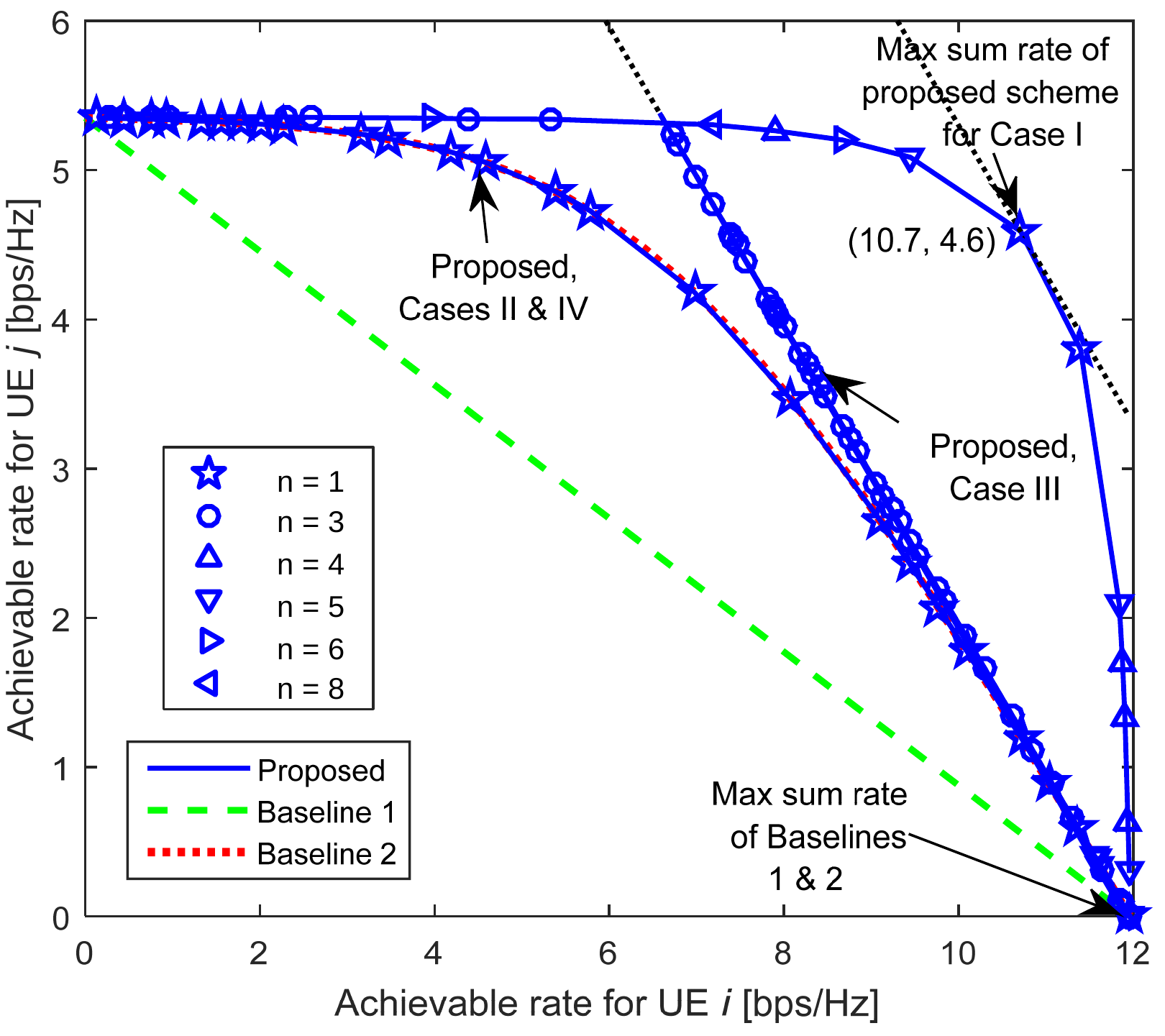}}
\caption{\label{fig3} Achievable rate region of the proposed scheme and Baselines 1 and 2 for (a) $\alpha_i = 10^{-3}$, $\alpha_j = 10^{-2}$ and (b) $\alpha_i = 10^{-3}$, $\alpha_j = 10^{-1}$, respectively. The rate tuple achieving the maximum sum rate is the intersection of the rate region boundary and the straight dotted lines with slope $-1$. For the proposed scheme, the decoding orders achieving the Pareto optimal boundary points are indicated by markers.}
\vspace{-.5cm} 
\end{figure}

In Figs.~\ref{fig3a} and \ref{fig3b}, we compare the achievable rate regions of the proposed cache-aided NOMA scheme and Baselines 1 and 2 for fixed channels with $\alpha_i = 10^{-3}$, $\alpha_j = 10^{-2}$ and $\alpha_i = 10^{-3}$, $\alpha_j = 10^{-1}$, respectively. For the proposed scheme, the rate achieved by UE $k$ is given by $r_{k,1} + r_{k,2}$, $k \in \{i, j\}$. Note that the achievable rate regions of Baselines 1 and 2 are independent of the values of caching vector $\mathbf{c}$. Hence, Baseline 2 with and without caching achieves the same rate region. From Figs.~\ref{fig3a} and \ref{fig3b} we observe that all considered schemes achieve the same corner points $\{(0, 8.7), (12.0,0)\}$ and $\{(0,5.3), (12.0,0)\}$, since the maximal rate for each UE is fundamentally limited by its channel status. Baseline 1 achieves the smallest rate region as it employs OMA to avoid interference. As NOMA introduces additional degrees of freedom for the users, Baseline 2 offers a larger achievable rate region than Baseline 1. The expansion of the rate region is more significant for the weak UE than for the strong UE, particularly when the differences of the UEs' channel gains are large, since the strong UE consumes a small transmit power, and hence, causes insignificant interference to the weak UE. However, as expected, Baselines 1 and 2 achieve the same maximum sum rate, i.e., maximum system throughput, of $12.0$ bps/Hz when only the strong UE transmits, i.e., no transmission time and power are allocated to the weak UE.

On the other hand, according to Corollary~\ref{cor1}, the achievable rate region of the proposed scheme depends on the cache configuration, although it is independent of $\mathbf{c}$ for a given cache configuration. For Cases \mbox{II} and \mbox{IV}, the achievable rate regions of the proposed cache-aided NOMA scheme and Baseline 2 coincide. This result is obvious for Case \mbox{IV}. For Case \mbox{II}, however, although CIC is enabled at the strong UE, for the considered parameter setting, the combination of CIC and SIC does not yield a gain compared to employing only SIC for the delivery of the corresponding subfiles. The proposed cache-aided NOMA scheme achieves larger rate regions for Cases \mbox{I} and \mbox{III}, where joint CIC and SIC enable the cancellation of more interference compared to Baseline 2 which performs only SIC. In particular, for Case \mbox{III}, CIC is enabled at the weak UE, which is not possible with conventional NOMA and improves user fairness. For Case \mbox{I}, joint CIC and SIC is enabled at both the strong and the weak UEs and CIC makes several SIC decoding orders feasible. Hence, in this case, the proposed scheme achieves the largest rate region among all considered schemes and significant performance gains for both the weak and the strong UEs are possible. Interestingly, different from the baseline schemes, for Cases~\mbox{I} and \mbox{III} of the proposed scheme, the maximum sum rate is achieved if the weak and the strong UEs receive data simultaneously. For Case \mbox{I}, simultaneous delivery to both users leads to a much higher system throughput (i.e., $18.6$~bps/Hz and $15.3$~bps/Hz in Figs. \ref{fig3a} and \ref{fig3b}, respectively) compared to Baselines 1 and 2 (i.e., $12.0$~bps/Hz). 

\begin{figure}[t]
\centering
\subfloat[]{\label{fig6a} \includegraphics[width=3.0in]{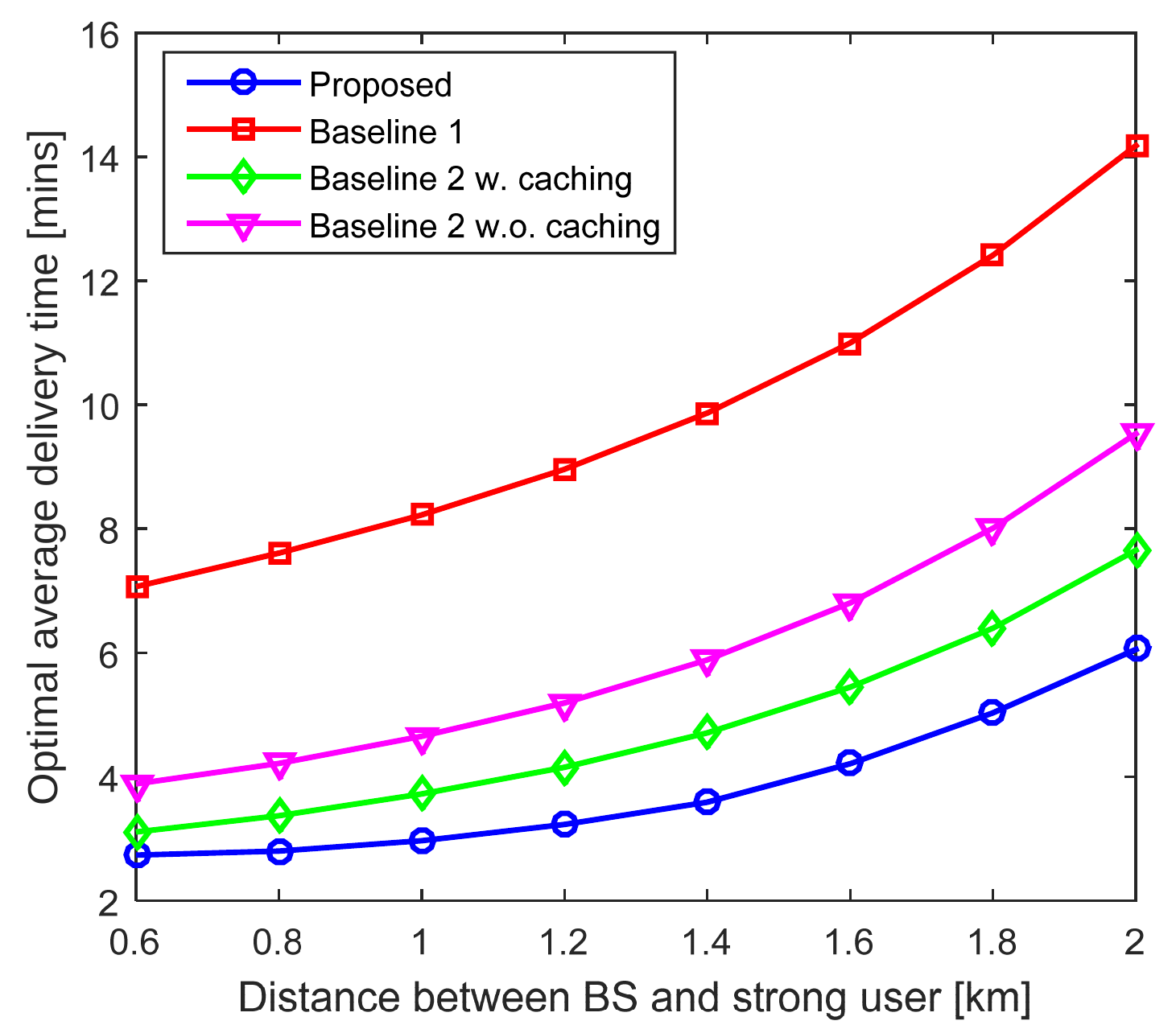}}  

\vspace{-.3cm}
\subfloat[]{\label{fig6b} \includegraphics[width=3.0in]{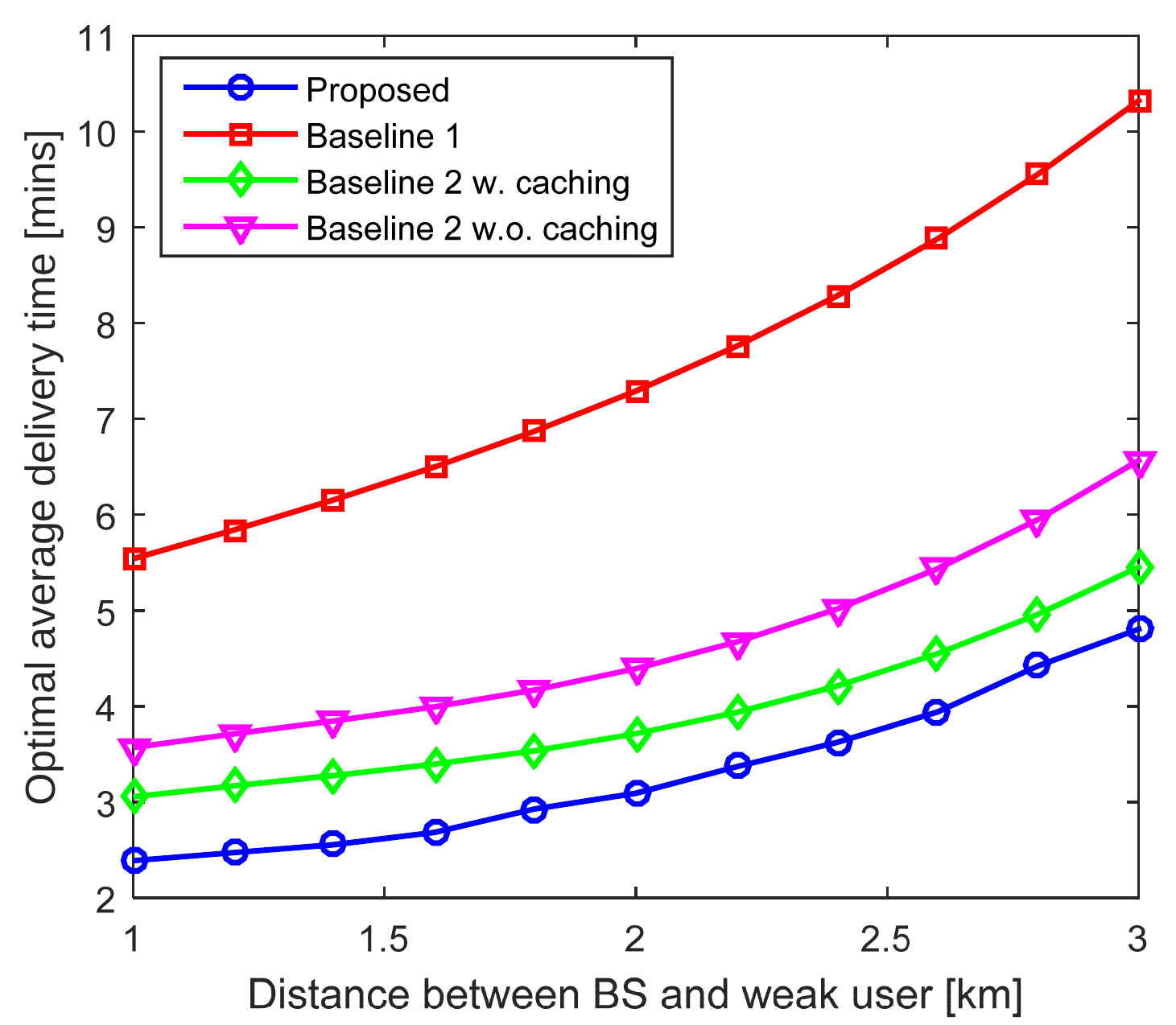}}
\caption{ Optimal average delivery time versus distance between the BS and (a) the strong user, $R_i$, and (b) the weak user, $R_j$.}
\vspace{-.5cm} 
\end{figure}

In the following, we adopt Rayleigh fading channels as outlined at the beginning of this section. All results are averaged over $100$ realizations of the UE locations and the channel fading. Figs.~\ref{fig6a} and \ref{fig6b} show the optimal average delivery times of the proposed cache-aided NOMA scheme and Baselines 1 and 2 as functions of the distances between the BS and the strong UE and between the BS and the weak UE, $R_i$ and $R_j$, respectively. As expected from the achievable rate regions in Fig.~\ref{fig3}, Baseline 1 requires the longest time to complete video file delivery. The proposed cache-aided NOMA scheme outperforms both Baseline 2 without caching and Baseline 2 with caching. This is due to the exploitation of  CIC, which is possible only with the proposed joint caching and NOMA transmission design. As $R_i$ and $R_j$ increase, the strong and the weak UEs suffer from increased path losses, which in turn reduces the channel gains of UE $j$ and UE $i$, respectively. Hence, we observe from Figs.~\ref{fig6a} and \ref{fig6b} that the optimal delivery time increases with $R_i$ and $R_j$ for all considered schemes. However, Baseline 1 is the least efficient among the considered schemes, and its delivery time almost doubles as $R_i$ and $R_j$ increase in the considered ranges. By exploiting the degrees of freedom offered by NOMA, Baseline 2 effectively reduces the performance degradation caused by the weak UE. For example, even without caching, the delivery time of Baseline 2 is $45$\% lower than that of Baseline 1 when the UEs are located at $R_i = 1$~km and $R_j = 2$~km, respectively. Moreover, when a cache is available, Baseline 2 can exploit caching for offloading the delivery data, which further reduces the delivery time compared to Baseline 1. The proposed scheme enjoys the best performance and its delivery time is at least $30$\% lower than that of Baseline 2 without caching for the considered values of $R_j$. The performance gap between cache-aided NOMA and Baseline 2 with caching grows as $R_i$ increases and $R_j$ decreases, since the proposed scheme reduces the performance degradation caused by the weak user. 

\begin{figure}[t]
\centering
\subfloat[]{\label{fig4a} \includegraphics[width=3.0in]{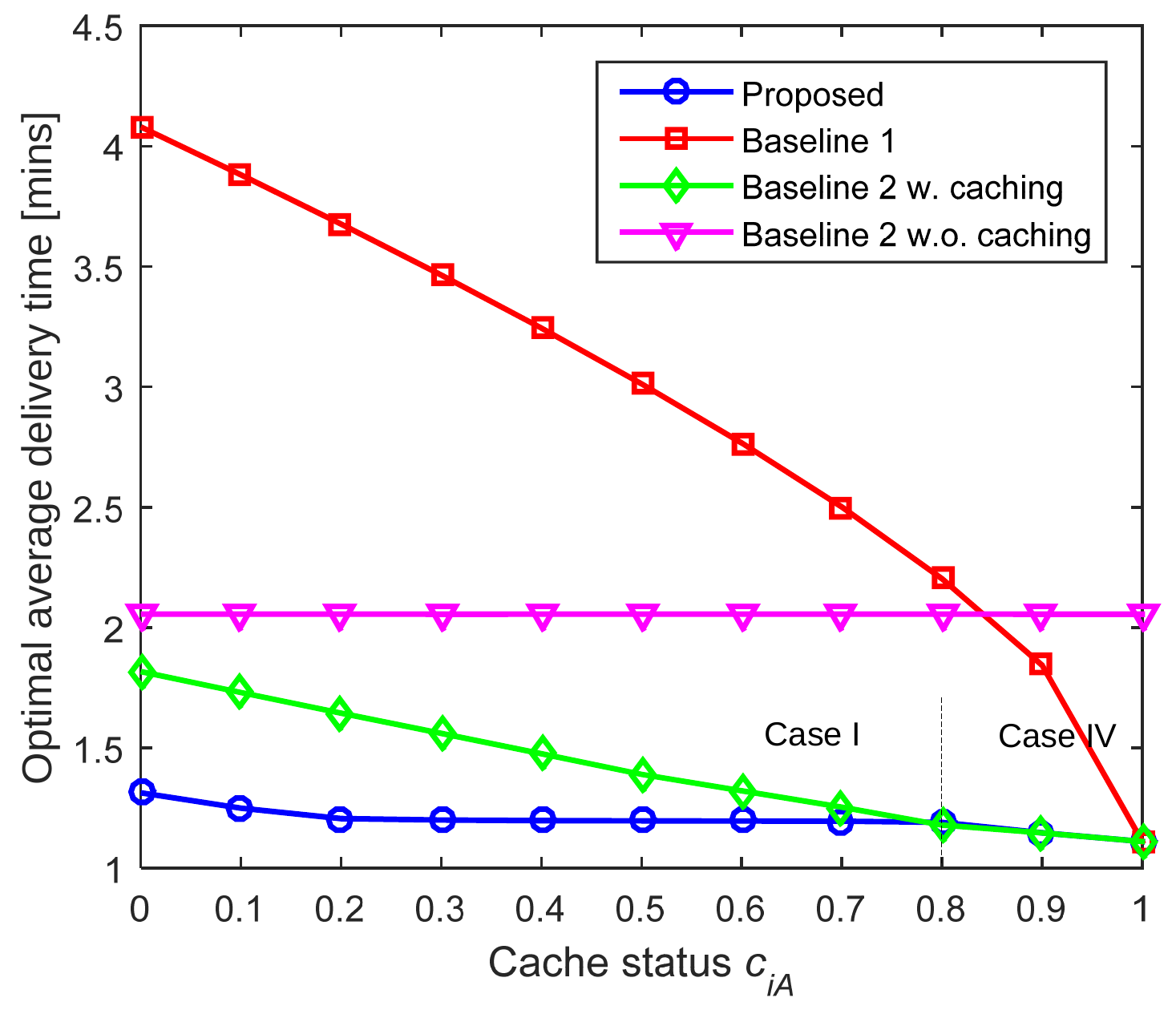}} 

\vspace{-.3cm}
\subfloat[]{\label{fig4b}  \includegraphics[width=3.0in]{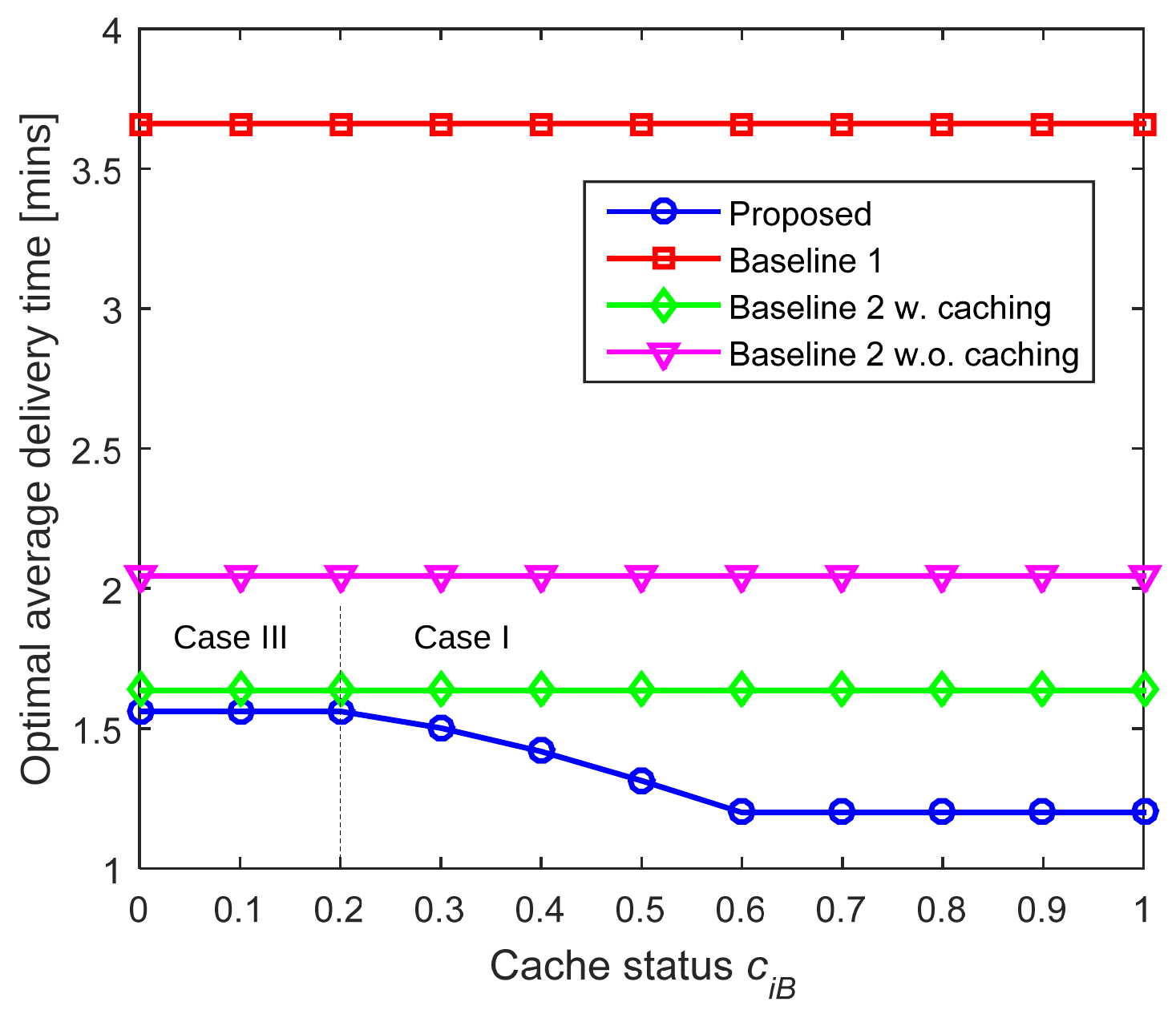} }
\caption{Optimal average delivery time versus cache status (a) $c_{iA}$ and (b) $c_{iB}$ for $R_i = 0.4$ km and $R_j = 0.6$ km.}
\vspace{-.5cm} 
\end{figure}

Figs.~\ref{fig4a} and~\ref{fig4b} show the optimal average delivery time of the considered schemes as functions of  $c_{iA}$ and $c_{iB}$, respectively, for $R_i = 0.4$~km and $R_j = 0.6$~km. Different from the achievable rates, the delivery times of Baseline 1, Baseline 2 with caching, and the proposed scheme critically depend on the amount of cached data. In particular, as $c_{iA}$ increases, the cached data is hit at UE $i$ with a higher probability; hence, more data is offloaded and the delivery times of Baseline 1, Baseline 2 with caching, and the proposed scheme decrease, cf. Fig.~\ref{fig4a}. When the requested file $W_A$ is completely cached at UE $i$, i.e., $c_{iA}= 1$, the delivery times of these three cache based schemes coincide. On the other hand, Fig.~\ref{fig4b} shows that, as $c_{iB}$ increases, only the proposed scheme can benefit from the cached data of the non-requested file $W_B$ at UE $i$ via CIC. As $c_{iA}$ and $c_{iB}$ change, different cache configurations become relevant for the proposed scheme. In particular, for given $c_{i,B}$, Cases~\mbox{I} and \mbox{IV} apply when small and large portions of the requested file $W_{A}$ are cached for offloading, respectively. In contrast, for given $c_{i,A}$, Cases~\mbox{III} and \mbox{I} apply when small and large portions of the non-requested file $W_B$ are cached for CIC, respectively.

\begin{figure}[t]
\centering
\subfloat[]{\label{fig5a} \includegraphics[width=3.0in]{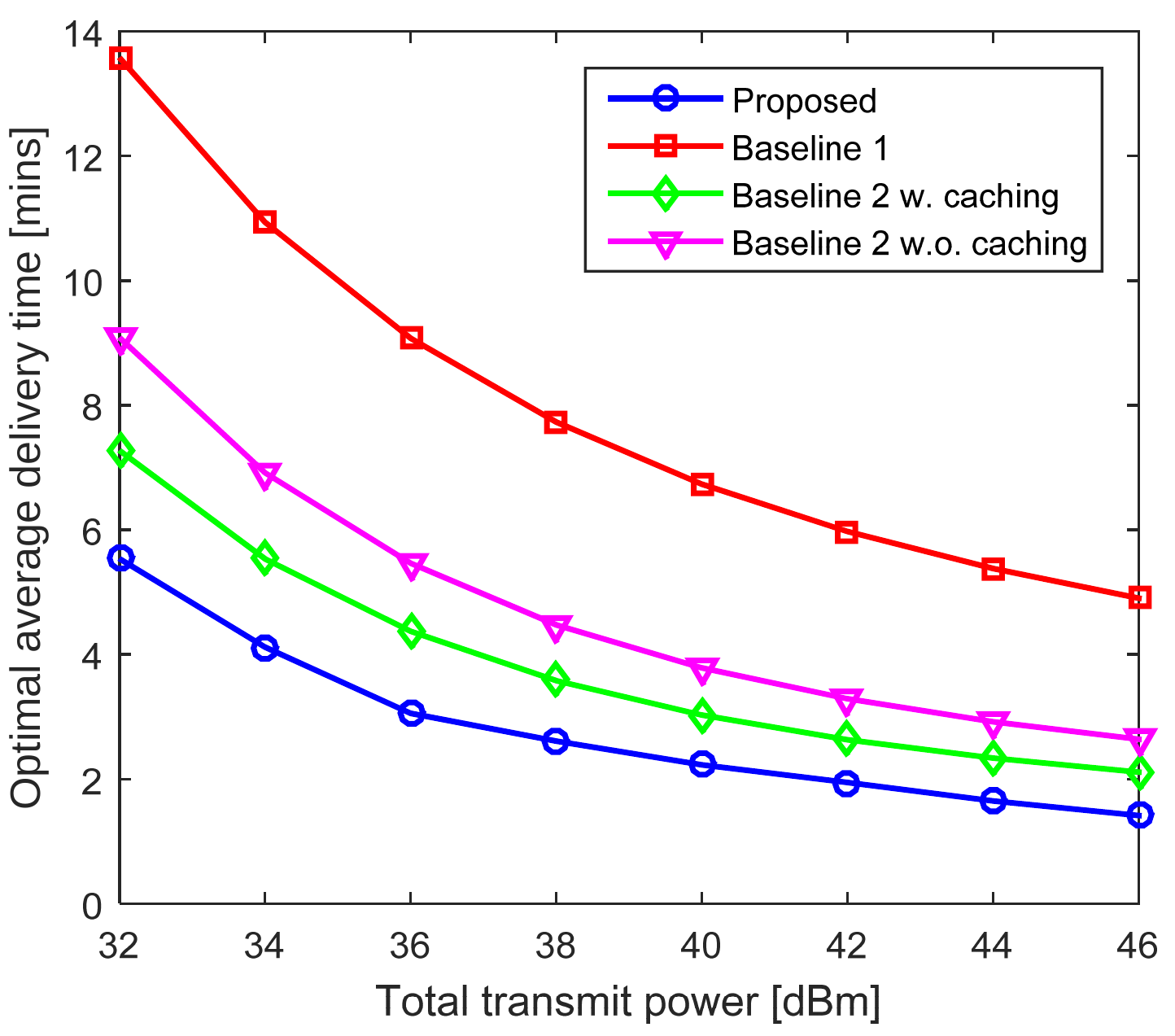}}  

\vspace{-.3cm}
\subfloat[]{\label{fig5b} \includegraphics[width=3.0in]{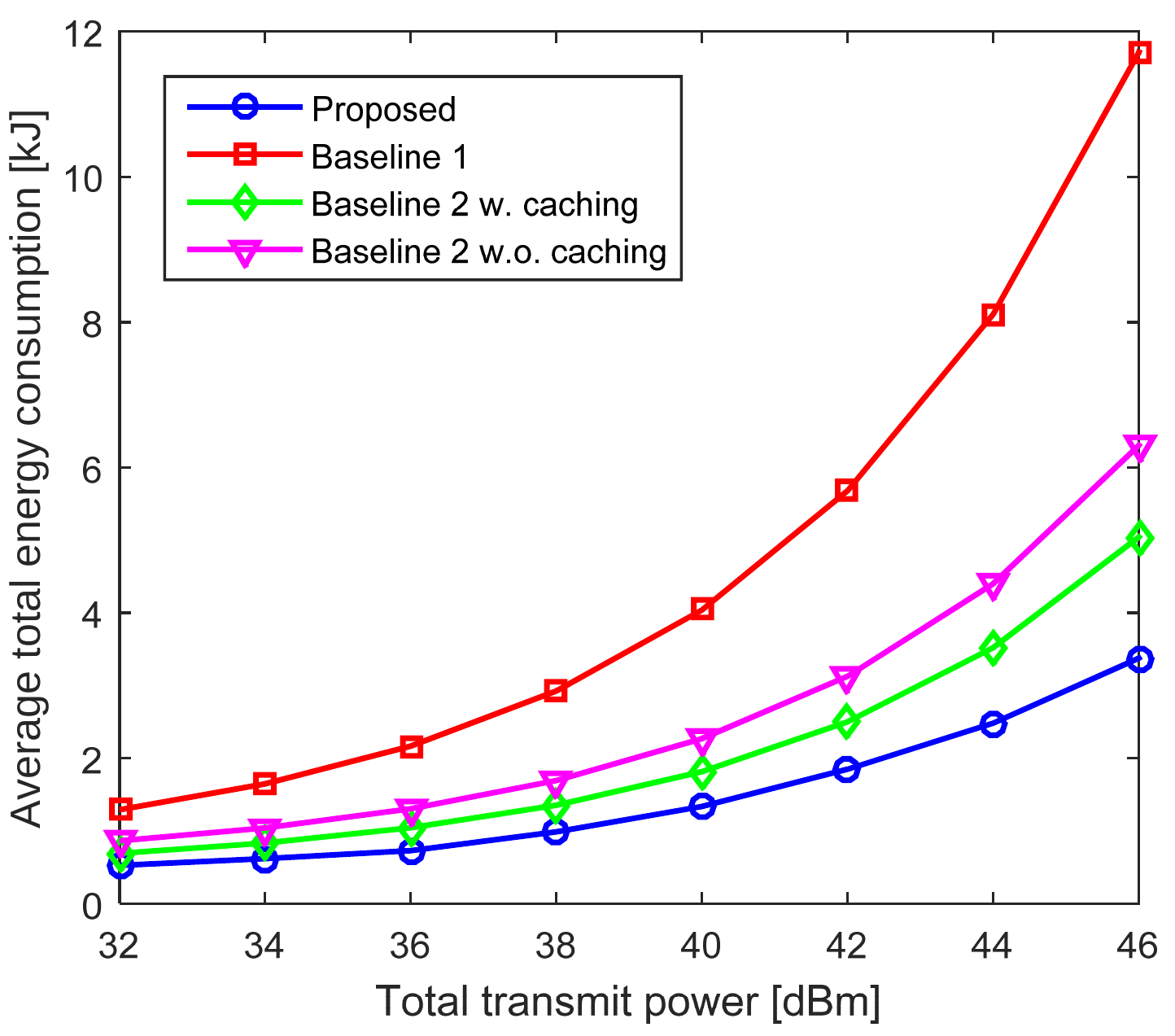}}
\caption{ (a) Optimal average delivery time and (b) average total transmission energy consumption versus total transmit power.}
\vspace{-.5cm} 
\end{figure}

Finally, Figs.~\ref{fig5a} and \ref{fig5b} show the optimal delivery time and the total transmission energy consumption as functions of the total transmit power, where the transmission energy consumption is defined as the product of transmit power and delivery time. For the low transmit power regime, the optimal average delivery time of all considered schemes decreases substantially as the total transmit power increases, such that the total energy consumption increases only moderately. By contrast, for the high transmit power regime, the additional delivery time reductions achieved by further transmit power increases saturate, whereby the total transmission energy consumption rapidly increases. Hence, there is an interesting trade-off between transmission energy consumption and delivery time for all the considered schemes. However, the proposed scheme outperforms all baseline schemes w.r.t. both delivery time and transmission energy consumption.

\section{\label{sec6}Conclusions}
In this paper, cache-aided NOMA was proposed for spectrally efficient downlink communication. The proposed scheme exploits unrequested cached data for cancellation of NOMA interference. This CIC is not possible with the conventional separate design of caching and NOMA. The achievable rate region of the proposed cache-aided NOMA scheme was characterized, and the optimal decoding order and power and rate allocations for minimization of the delivery time were derived in closed form. Simulation results showed that the proposed scheme can significantly enlarge the achievable downlink rate region for both the strong and the weak users. Moreover, unlike conventional NOMA, cache-aided NOMA can not only improve user fairness but can also significantly increase the achievable sum rate compared to OMA by enabling joint CIC and SIC. Furthermore, the delivery time of both users was shown to be substantially reduced compared to OMA and conventional NOMA with and without caching. 

{{In this paper, we only considered the delivery problem in a single-cell network with single-antenna BS and UEs. Extending the proposed cache-aided NOMA scheme to multi-cell multi-antenna networks and optimizing the caching phase are interesting topics for future research. Moreover, the impact of stochastic traffic loads and network (user/BS) deployments on the performance of cache-aided NOMA requires further study.}}

\vspace{-.2cm}
\appendix[Proof of Proposition \ref{prop1}]
As $i\overset{_{(1)}}{\to}x_{A1}$, $r_{i,1}\le C\left(\tfrac{p_{i,1}}{p_{i,2}+p_{j,2}+\alpha_{i}}\right)$ is achievable for decoding $x_{A1}$ at UE $i$. To derive the achievable rate region, we need to check the decodability of the interfering signals $x_{B2}$ and $x_{A2}$ at UE $i$ and $j$, respectively. Let us consider the following two power regions. 

\subsubsection{Region 1} For $\mathbf{p}\in \mathcal{P}_{\lyxmathsym{\mbox{I}},1} \backslash \mathcal{P}_{\lyxmathsym{\mbox{I}},2}$, we have $C\left(\tfrac{p_{i,2}}{p_{j,1}+\alpha_{j}}\right)<C\left(\tfrac{p_{i,2}}{p_{j,2}+\alpha_{i}}\right)$, i.e., UE $j$ cannot decode $x_{A2}$ before decoding $x_{B1}$ as the SIC decoding condition is not met. Also, for any $\mathbf{p} \in\mathbb{R}_{+}^{4}$, UE $j$ cannot decode $x_{A2}$  before decoding $x_{B2}$ as
\begin{equation}
C  \left( \tfrac{p_{i,2}}{p_{j,1}+p_{j,2}+\alpha_{j}}  \right)  <  C  \left( \tfrac{p_{i,2}}{p_{j,2}+\alpha_{j}}  \right)  <  C  \left( \tfrac{p_{i,2}}{p_{j,2}+\alpha_{i}}  \right) .
\label{eq:26-1}
\end{equation}
On the other hand, for any $\mathbf{p} \in\mathbb{R}_{+}^{4}$, $x_{B2}$ can be always decoded and canceled at UE $i$ before $x_{A2}$ is decoded as $\alpha_{i}<\alpha_{j}$; and hence, $r_{i,2}\le C\left(\frac{p_{i,2}}{\alpha_{i}}\right)$ is achievable. In contrast,  UE $j$ cannot decode $x_{A2}$ in any case. Consequently, the feasible decoding orders are UE $i\overset{_{(2)}}{\to}x_{B2}\overset{_{(3)}}{\to}x_{A2}$ and UE $j\overset{_{(1)}}{\to}(x_{B1}, x_{B2})$, whereby rate region $\mathcal{R}_{\lyxmathsym{\mbox{I}},1}\left( \mathcal{P}_{\lyxmathsym{\mbox{I}},1} \backslash \mathcal{P}_{\lyxmathsym{\mbox{I}},2} \right)$ is achieved.

\subsubsection{Region 2} For $\mathbf{p}\in\mathcal{P}_{\lyxmathsym{\mbox{I}},2}$, we have $C\left(\frac{p_{i,2}}{\alpha_{i}}\right)>C\left(\frac{p_{i,2}}{p_{j,1}+\alpha_{j}}\right)>C\left(\frac{p_{i,2}}{p_{j,2}+\alpha_{i}}\right)$, i.e., $x_{A2}$ can be decoded at UE $j$ before  $x_{B1}$ is decoded if and only if UE $i \overset{_{(2)}}{\to} x_{A2}$. Assume $x_{A2}$ is decoded last at UE $i$ such that UE $j$ cannot decode $x_{A2}$ in any case. Then, rate region $\mathcal{R}_{\lyxmathsym{\mbox{I}},1}\left(\mathcal{P}_{\lyxmathsym{\mbox{I}},2}\right)$ is achievable. On the other hand, assume UE $i \overset{_{(2)}}{\to} x_{A2}$. Then, UE $j$ can achieve a higher rate for $r_{j,1}$ by decoding $x_{A2}$ before decoding $x_{B1}$, which is only possible after $x_{B2}$ has been decoded according to \eqref{eq:26-1}. Thus, the rate region $\mathcal{R}_{\lyxmathsym{\mbox{I}},2}\left(\mathcal{P}_{\lyxmathsym{\mbox{I}},2}\right)$ is achievable.

Therefore, the rate region $\mathcal{R}_{\lyxmathsym{\mbox{I}},1}\left(\mathcal{P}_{\lyxmathsym{\mbox{I}},1}\right)\bigcup\mathcal{R}_{\lyxmathsym{\mbox{I}},2}\left(\mathcal{P}_{\lyxmathsym{\mbox{I}},2}\right)$ is achievable, and any rate vector outside the region $\mathcal{R}_{\lyxmathsym{\mbox{I}},1}\left(\mathcal{P}_{\lyxmathsym{\mbox{I}},1}\right)\bigcup\mathcal{R}_{\lyxmathsym{\mbox{I}},2}\left(\mathcal{P}_{\lyxmathsym{\mbox{I}},2}\right)$ cannot be achieved by SIC decoding. This completes the proof.

\IEEEtriggeratref{23}

\end{document}